\documentclass[a4paper]{article}
\pdfoutput=1 

\usepackage{jheppub} 

\usepackage{graphicx,here,epsfig,wrapfig,amssymb,color,amsmath,bm,breqn}

\usepackage[T1]{fontenc} 
\usepackage{multirow}
\usepackage{amsmath}
\usepackage{pifont}
\usepackage{float}
\usepackage{subfig}
\allowdisplaybreaks
\usepackage{soul}
\usepackage[utf8]{inputenc} 
\usepackage{cancel}
\usepackage{multirow,boldline}

\usepackage{lineno} 
\usepackage[export]{adjustbox}
\usepackage{hyperref} 
\hypersetup{
    colorlinks,
    citecolor=black,
    filecolor=black,
    linkcolor=black,
    urlcolor=black
}

\renewcommand{\(}{\left(}
\renewcommand{\)}{\right)}

\renewcommand{\vec}[1]{\bm{#1}}

\newcommand{\nn}{\nonumber}
\newcommand{\ot}{\leftarrow}

\newcommand{\specialcellcenter}[2][c]{\begin{tabular}[#1]{@{}c@{}}#2\end{tabular}}

\renewcommand{\(}{\left(}
\renewcommand{\)}{\right)}

\title{PDF bias and flavor dependence in TMD distributions}

\author[a]{Marcin Bury,}
\author[b,c,d]{Francesco Hautmann,}
\author[e]{Sergio Leal-Gomez,}
\author[f]{Ignazio Scimemi,}
\author[f,g]{Alexey Vladimirov,}
\author[g]{Pia Zurita}

\emailAdd{marcin.bury@ifj.edu.pl}
\emailAdd{francesco.hautmann@physics.ox.ac.uk}
\emailAdd{sergio.leal.gomez@univie.ac.at} 
\emailAdd{ignazios@ucm.es}
\emailAdd{alexey.vladimirov@ur.de}
\emailAdd{maria.zurita@ur.de}

\affiliation[a]{
Jerzy Haber Institute of Catalysis and Surface Chemistry, Polish Academy
of Sciences, Niezapominajek 8, 30-239 Krak\'ow, Poland}
\affiliation[b]{CERN, Theory Department, CH 1211 Geneva, Switzerland}
\affiliation[c]{Elementaire Deeltjes Fysica, Universiteit Antwerpen, B 2020 Antwerpen, Belgium}
\affiliation[d]{
Theoretical Physics Department, University of Oxford, Oxford OX1 3PU, UK}
\affiliation[e]{
Wien Universit\"at, Faculty of Physics, Boltzmanngasse 5, A-1090 Vienna, Austria}
\affiliation[f]{Dpto. de F\'{i}sica Te\'{o}rica \& IPARCOS, Universidad Complutense de Madrid, E-28040 Madrid, Spain}
\affiliation[g]{Institut f\"ur Theoretische Physik, Universit\"at Regensburg, D-93040 Regensburg, Germany}

\preprint{CERN-TH-2022-126 \;\;\;\;  UWThPh 2021-29}
\abstract{
Transverse momentum dependent (TMD) distributions   
match collinear parton density functions (PDF)  in the limit of small transverse distances,  
 which is accounted for by global extractions of TMD distributions.  
We study  the influence of the collinear PDF value and uncertainties on the determination of unpolarized TMD distributions and the description of  Drell-Yan (DY)  and Z-boson production measurements at low transverse momenta. 
We take into account,  for the first time, flavor-dependent  
non-perturbative TMD profiles.  
We carry out a Bayesian analysis to incorporate the 
propagation of PDF uncertainties into TMD extractions.
We find that collinear PDF uncertainties and 
non-perturbative TMD flavor dependence are both 
essential to obtain reliable TMD determinations,  
and should be included in future global analyses.
}

\date{\today}

\begin{document}

\maketitle

\section{Introduction}

While determinations of the proton's collinear parton distribution functions (PDFs) \cite{Kovarik:2019xvh} from fits to experimental data have been carried out for four decades, and PDFs are nowadays known with high precision, the systematic investigation of the transverse momentum dependent (TMD) \cite{Angeles-Martinez:2015sea}  distribution functions has started much more recently.  In the last few years determinations of TMD distributions \cite{Aybat:2011zv} from fits to experimental data have been performed in the context of low transverse momentum Drell-Yan (DY) production and semi-inclusive deep inelastic scattering (SIDIS) \cite{Scimemi:2019cmh,Bertone:2019nxa,Scimemi:2017etj,Bacchetta:2019sam,Bacchetta:2017gcc}, small-$x$ deep inelastic scattering \cite{Hautmann:2013tba},  nonlinear TMD evolution \cite{Kotko:2017oxg}, and parton branching   \cite{BermudezMartinez:2018fsv}. The results are collected in the public library  TMDlib~\cite{Abdulov:2021ivr,Hautmann:2014kza}, similarly to LHAPDF for the case of PDF.


A physical cross section $\sigma$ with  hard scale $Q$ and measured transverse momentum $q_T \ll Q$  is described  in terms of TMD distributions by a factorization formula with  the schematic form (up to power corrections in $q_T/Q$ and $\Lambda_{\rm{QCD}}/Q$) 
\begin{equation} 
\label{partonfact} 
{{d \sigma} \over { d q_T^2} } =  \sum_{i,j}  \int d^2 b \ e^{i b  \cdot q_T} \sigma^{(0)}_{i j} f_{1,i\leftarrow h1}(x_1,b;\mu,\zeta_1)f_{1,j\leftarrow h2}(x_2,b;\mu,\zeta_2) , %
\end{equation} 
where $b$ is the transverse distance Fourier conjugate to transverse momentum, $ \sigma^{(0)}_{i j}$ are perturbatively calculable partonic cross sections, and  $f_{i}$ and $f_{j}$ are  TMD parton distributions. 
These depend on the mass and rapidity scales $\mu$ and $\zeta$ through appropriate evolution equations, involving both perturbative and non-perturbative (NP)  components, and are to be  determined from fits to experiment.

Current fits of TMD distributions from DY and SIDIS measurements are performed by using not only the TMD factorization and evolution framework but also  
additional inputs which exploit the relationship between TMD distributions and the collinear PDFs. These relations follow from the operator product expansion (OPE) applied 
to the TMD operator. 
The OPE expresses the TMD distributions 
at small distances $b$   
in terms of PDFs via perturbatively calculable  
coefficients, 
 with power corrections in $b$,  
as follows 
\begin{equation}\label{def:F1=C.f1}
f_{1,f\ot h}(x,b;\mu,\zeta)=\sum_{f'}\int_{x}^1 \frac{dy}{y}C_{f\ot f'}\(y,b;\mu,\zeta\)q_{f'}\(\frac{x}{y},\mu\)+O(b^2),
\end{equation}
where $f_i(x,b;\mu,\zeta)$ is the  TMD distribution, $q_j(x,\mu)$ is the  PDF, $C_{i j}$ are the Wilson coefficient functions, $\mu$ and $\zeta$ are factorization scales, the subscripts $f$ and $f'$ indicate  parton flavors, and the last term on the right hand side is the power suppressed correction at small $b$. 
An expansion such as 
eq.~(\ref{def:F1=C.f1}) is used, for instance,   
 in the context of precision studies of the DY transverse momentum $q_T$ spectrum at the Large Hadron Collider (LHC) \cite{Lhcew:2022pt}   to carry out  comparisons of small-$q_T$ logarithmic resummations between  computer codes     
 based on TMD distributions (e.g., 
 \texttt{artemide} \cite{Scimemi:2017etj}, \texttt{NangaParbat}  \cite{Bacchetta:2019sam})   and computer codes 
 which perform perturbative small-$q_T$ resummation without systematically introducing TMD distributions   
 (e.g., 
 \texttt{DYturbo}   \cite{Camarda:2019zyx,Camarda:2021ict},  \texttt{reSolve}   \cite{Coradeschi:2017zzw,Accomando:2019ahs}, \texttt{Radish}   \cite{Bizon:2018foh,Chen:2022cgv}, \texttt{SCETlib}   \cite{Ebert:2020dfc,Ebert:2016gcn},  \texttt{CuTe}   \cite{Becher:2020ugp}).  
%

To carry out TMD fits an ansatz is made 
for the  NP TMD distribution at large 
$b$, where  the power corrections to OPE become sizeable.  
This is of the form 
\cite{Bacchetta:2017gcc,Scimemi:2017etj,Bacchetta:2019sam,Bertone:2019nxa,Scimemi:2019cmh,Vladimirov:2019bfa,Bury:2020vhj,Echevarria:2020hpy} 
\begin{equation}\label{def:ansatz-general}
f_{1,f\ot h}(x,b;\mu,\zeta)
=\sum_{f'}\int_{x}^1 \frac{dy}{y}C_{f\ot f'}\(y,b;\mu,\zeta\)q_{f'}\(\frac{x}{y},\mu\) f_{\text{NP}}^f(x,b) ,  
\end{equation}
where $f_{\text{NP}}^f$ is a function which contains power corrections to eq.~(\ref{def:F1=C.f1}), behaves as $f_{\text{NP}}^f(x,b)\sim 1+O(b^2)$ for $b\to 0$, and is to be fitted to experimental data. This ansatz  reproduces  eq.~(\ref{def:F1=C.f1}) for $b\to 0$ but allows for modifications at finite $b$. Current fits determine the NP TMD distributions, once  a choice is made for the collinear PDFs in eq.~(\ref{def:ansatz-general}).

The PDF choice in eqn.(\ref{def:ansatz-general}) determines the value of TMD distribution and influences $f_{\text{NP}}$ (we address this as {\it PDF bias}). Consequently, the extraction based solely on the central value of PDF are incomplete, and PDF uncertainties should be incorporated. The way the PDF error propagates  has nontrivial consequences 
which have  not been studied in the literature so far.  The purpose of this work is to 
develop methods for addressing these issues, to 
critically assess 
the current status of  phenomenological analyses based on TMD factorization 
  and OPE expansion, and to   perform the first systematic 
investigation of the role of PDF bias in TMD determinations.   

To this end, we take PDF sets  
\texttt{HERA20}~\cite{Abramowicz:2015mha}, 
 \texttt{NNPDF3.1}~\cite{Ball:2017nwa}, 
\texttt{CT18}~\cite{Hou:2019efy}  and 
\texttt{MSHT20}~\cite{Bailey:2020ooq}
as representatives of different methodological approaches at 
next-to-next-to-leading order (NNLO); we 
use the implementation of the TMD factorization 
formula as in refs.~\cite{Scimemi:2019cmh,Hautmann:2020cyp} to 
 perform fits of  experimental measurements for 
unpolarized DY and $Z$-boson production at small $q_T$, from 
fixed-target to LHC energies; we devise and implement a Bayesian 
procedure to propagate the PDF uncertainties, for each PDF set, to 
TMD extractions. 
This enables us to present results for the fitted NP TMD distributions 
which  display, for the first time, both experimental 
and PDF uncertainties. We find that the role of the collinear PDFs is very significant in current TMD phenomenology. 
We discuss the issues associated with the 
combination of the different sources of TMD uncertainty, 
and propose an approach based on  the bootstrapping method.

 %


An important  aspect addressed by our work concerns the physical 
properties of the NP distributions $f_{\text{NP}}^f$  in 
eq.~(\ref{def:ansatz-general}).  The fits  based on TMD 
factorization and evolution which have so far been performed in the 
literature  have assumed flavor-independent 
parameterizations for $f_{\text{NP}}^f$, with the 
flavor dependence solely in collinear PDFs, 
motivated by the belief that 
available data have little sensitivity to TMD flavor structure. In this work 
we go beyond this assumption and include flavor dependence in 
the NP TMD distributions. We find that the 
 the inclusion of TMD flavor dependence improves the quality of the 
 fits significantly, leading  
  both to a better agreement of theory with data and to 
more consistent results among different collinear PDF sets. 
For each set, we examine the distribution of $\chi^2$ values 
over PDF replicas. We  demonstrate that 
the   spread in the $\chi^2$ distribution among replicas 
is much reduced  with flavor-dependent TMD 
   compared to the flavor-independent case, and that this 
   conclusion applies consistently across  collinear sets.

A complete treatment of the problems studied in this paper 
should eventually involve simultaneous fits of PDF and TMD distributions,  requiring much more computational power than is available to us at present. 
The analysis presented in this paper underlines essential elements which should be 
included in future TMD studies. 

The article is structured as follows. Sec.~\ref{sec:notation}  provides basic theoretical inputs and notation, and  
presents the  flavor dependent model used for $f_{\rm NP}^f$. 
 Sec.~\ref{sec:data}  describes the data used in the analysis. The statistical methods are reported in sec.~\ref{sec:statistical}. Sec.~\ref{sec:results} presents the results of the fits and the discussion of their consequences. We give conclusions in sec.~\ref{sec:conclusions}. We collect further details on the fits in the appendix.  
 
%


\section{Theory inputs}
\label{sec:notation}

In this section we summarize the basic theoretical inputs of our analysis:   
i) TMD factorization formula for the DY cross section and  
evolution equations for the TMD distributions, 
and ii) models for the non-perturbative contributions  to TMD 
factorization and evolution formulas. 

The new element in this section is the flavor-dependent TMD profile in 
eq.~(\ref{def:fNP}): to our knowledge, 
flavor dependence of the TMD has not been considered 
before in the context of DY or SIDIS fits  including TMD evolution.

\subsection{DY factorization formula}
\label{app:DY}

The TMD factorized expression for the differential 
cross-section of the DY process $h_1+h_2\to \gamma^*/Z(\to \ell^+\ell^-)+X$ in the low transverse momentum region 
can be written as 
\cite{Collins:2011zzd,Collins:1984kg} 
\begin{eqnarray}\label{DY:xSection}
\frac{d\sigma}{dQ^2dy d\vec q_T^2}&=&
 \sum_{f} \sigma_0   
\int_0^\infty db\,bJ_0(bq_T)  H ( Q , \mu) 
 f_{1,f\ot h_1}(x_1,b;\mu,\zeta_1)f_{1,\bar f\ot h_2}(x_2,b;\mu,\zeta_2)   
 \\\nn && \qquad    
+ {\cal O} \left( {q_T / Q} \right) + {\cal O} \left( {\Lambda_{\rm{QCD}} / Q} \right)  . 
\end{eqnarray}
Here $Q^2$ is the invariant mass of the vector boson, $\vec q_T$ is the transverse component of its momentum  (relative to the scattering plane), and 
$\Lambda_{\rm{QCD}} $ is the characteristic  hadronic scale. 
The index $f$ runs over all active quark flavors. The distribution 
$f_1$ is the unpolarized TMDPDF, 
depending on the lightcone momentum fraction $x$,   
transverse distance $b$, mass and rapidity 
scales $\mu$ and $\zeta$; $\sigma_0$ is the 
leading-order DY cross section, $H$ is the perturbative hard factor 
containing higher-order corrections,  and $J_0$ is the Bessel 
function.  
The formula is valid up to power-suppressed corrections in 
$ q_T / Q $ and $  \Lambda_{\rm{QCD}} / Q $,   indicated in the 
second line of eq.~(\ref{DY:xSection}). 
The DY factorization formula (\ref{DY:xSection})  has been rederived 
in 
 \cite{Becher:2010tm,Echevarria:2011epo,Chiu:2012ir}.  We use 
 its implementation as given in 
 refs.~\cite{Scimemi:2017etj,Scimemi:2018xaf,Scimemi:2019cmh,Vladimirov:2019bfa,Bertone:2019nxa,Gutierrez-Reyes:2019vbx,Hautmann:2020cyp,Gutierrez-Reyes:2020ouu,Bury:2020vhj,Moos:2020wvd,Vladimirov:2021hdn}.

The evolution of TMD distributions  in mass and rapidity  is given by the pair of equations
\begin{align}\label{evoequ}
\mu^2\frac{\partial f_{1,f\ot h}(x,b; \mu,\zeta)}{\partial \mu^2}&=
\(\Gamma_{\rm cusp}\ln\frac{\mu^2}{\zeta}-\gamma_V(\mu)\) f_{1,f\ot h}(x,b; \mu,\zeta) ,\nn \\
\zeta\frac{\partial f_{1,f\ot h}(x,b; \mu,\zeta)}{\partial \zeta}&=-{\cal D}(b,\mu)f_{1,f\ot h}(x,b; \mu,\zeta),
\end{align}
where $\Gamma_{\rm cusp}$ is the anomalous dimension for the cusp of light-like Wilson lines, $\gamma_V$ is the anomalous dimension of the quark vector form factor and $\mathcal{D}$ is the  Collins-Soper (CS) kernel~\cite{Collins:1981uk,Collins:1981va}. 
We will use the solution of the evolution equations  
 according to the 
$\zeta$-prescription \cite{Scimemi:2017etj,Scimemi:2018xaf,Vladimirov:2019bfa,Scimemi:2019cmh}. Using the expansion (\ref{def:ansatz-general}), this solution 
is expressed in terms of PDFs,  
$f_{\text{NP}}^f$ and 
the matching coefficients \cite{Echevarria:2016scs,  
Luo:2020epw,Luo:2019szz,Ebert:2020yqt}.  


The perturbative orders in the strong coupling $\alpha_s$ which will be used in the calculations  presented in the following sections 
are summarised in tab.~\ref{tab:resume} for each element of the cross section.   
The definition of $\mathcal{D}_{\text{resum}}$ is given in the next subsection. The resulting cross section corresponds to 
the logarithmic order NNLL$^\prime$  according to the 
terminology in  \cite{Lhcew:2022pt}. 

\begin{table}[h]
\centering
\begin{tabular}{||c|c|c|c|c|c||}
\hline
$H$  &  $C_{f\leftarrow f'}$&$\Gamma_{\rm cusp}$& $\gamma_V$ & $\mathcal{D}_{\text{resum}}$ & $\alpha_s$ running $\&$ PDF evolution\\
\hline
$\alpha_s^2$   &  $\alpha_s^2$&  $\alpha_s^3$& $\alpha_s^2$ &  $\alpha_s^2$& NNLO\\
\hline
\end{tabular}
\caption{Summary of the perturbative orders used for each element of the cross section.\label{tab:resume}}
\end{table}



\subsection{Non-perturbative models}
\label{subsec:nonpertinputs}

The 
NP content of the DY differential cross section is encoded in two  functions: the CS kernel 
$\mathcal{D}$   in eq.~(\ref{evoequ})   
and the  TMD distribution 
$f_{\text{NP}}^f$   in eq.~(\ref{def:ansatz-general}).  

We write the  CS kernel as 
\begin{eqnarray}\label{def:CS-kernel}
\mathcal{D}(b,\mu)=\mathcal{D}_{\text{resum}}(b^{*}(b),\mu)+c_{0}bb^{*}(b) \, ,
\end{eqnarray}
where  the first term  $\mathcal{D}_{\text{resum}}$ on the right hand side 
is the resummed expression for $\mathcal{D}$ at small  $b$ \cite{Echevarria:2012pw} with NNLO perturbative coefficients \cite{Echevarria:2015byo,Vladimirov:2016dll}, 
while the second term is the non-perturbative model, given in 
terms  of 
$ b^{*}(b)=b / \sqrt{1+(b/B_{\rm NP})^2} $, with parameters  $c_0$ and 
$B_{\rm NP}$ to be determined from data. 
Motivated by the results  
of the fit \cite{Scimemi:2019cmh}, 
for simplicity in what follows we will take 
a fixed value $B_{\rm NP} = 2$ GeV$^{-1}$ 
and leave only  $c_0$ to be fitted to data.  
The model in eq.~(\ref{def:CS-kernel}) is linear at $b\to\infty$: this ansatz is partially supported by the studies \cite{Vladimirov:2020umg,Collins:2014jpa} and 
 lattice computations \cite{LatticeParton:2020uhz,Schlemmer:2021aij,Shanahan:2021tst}, 
and it has already been used in ref.~\cite{Scimemi:2019cmh}. 
In refs.~\cite{Hautmann:2020cyp,Hautmann:2021ovt} alternative models,  with 
quadratic asymptotic behavior (as in the 
studies \cite{Konychev:2005iy,Landry:2002ix,Landry:1999an}) 
and with constant asymptotic behavior (in the spirit of 
the $s$-channel 
picture \cite{Hautmann:2007cx,Hautmann:2000pw,Hautmann:1999ui}),   
have also been analyzed. Corresponding investigations based on these 
different asymptotic behaviors are left to future work.

The main new feature of the NP treatment 
in the present analysis concerns the TMD  
distribution $f_{\text{NP}}^f$. 
We include 
  flavor dependence in the TMD profile, and take the parameterization 
\begin{eqnarray}\label{def:fNP}
f_{\text{NP}}^{f}=\exp\(-\frac{(1-x)\lambda_1^f+x\lambda_2^f}{\sqrt{1+\lambda_0 x^2 b^2}}b^2\),
\end{eqnarray}
with $\lambda_{1,2}^f>0$ and $\lambda_0>0$. 
The model is characterized by an exponential asymptotic fall-off  at $b\to\infty$ and a Gaussian-like shape at intermediate $b$    
\cite{Landry:2002ix,Guzzi:2013aja,Schweitzer:2010tt,Schweitzer:2012hh,Scimemi:2016ffw,Scimemi:2017etj}. 
While the parameter $\lambda_0$ is taken to be 
universal for all flavors, 
the parameters $\lambda_{1,2}$ are taken  to be flavor dependent. We distinguish $u$, $d$, $\bar u$, $\bar d$ and $sea$ cases, where $sea$ is used for $(s, \bar s, c, \bar c, b ,\bar b)$  flavors. In total we have 11 free parameters. Since the parameters $\lambda_{1,2}$ are almost uncorrelated across flavors, the presence of unnecessary fitting parameters (i.e. not well restricted in a particular setup) does not lead to an overfit.
 
The numerical implementation is made with \texttt{artemide} \cite{Scimemi:2017etj}, which can be found in open-access repository \cite{web}. The PDF values and  their evolution 
are taken from  LHAPDF \cite{Buckley:2014ana}. \texttt{Artemide} is a  FORTRAN code, with a Python interface. The evaluation of a single DY-cross-section point includes two convolutions of PDFs, Hankel type integrals and three integrations over the phase space. Additionally, for some measurement one needs to take into account fiducial cuts. The \texttt{artemide} code is optimized for such computations, and it uses various numerical tricks, such as precomputed grids, specially optimized integration algorithms and parallel evaluation. Even so the computation of one value of $\chi^2$ for the data set takes from 30 seconds to a few minutes, depending on the PDF input, NP parameters,
experimental fiducial cuts and hardware. Therefore, a single minimisation procedure (made with the help of the iMINUIT package \cite{iminuit}) requires typically a few dozen hours on an average computer. 

\section{TMD data sets}
\label{sec:data}

\subsection{Complete data set}

The complete data set used in this study is given 
in tab.~\ref{tab:data}.    We restrict the  fit to data points in the low transverse momentum region by implementing 
the cut $q_T/Q<0.25$ \footnote{Treating the region $q_T \sim Q$ requires the inclusion of power corrections in $q_T/Q$ 
or matching with finite-order, NLO or higher, hard scattering 
coefficients \cite{Camarda:2019zyx,Bizon:2018foh,BermudezMartinez:2019anj,Becher:2020ugp} 
(see also \cite{BermudezMartinez:2020tys} for discussion of  different matching methods \cite{Collins:1984kg,Collins:2000gd}).
}. In addition  we implement an extra cutting rule for the very precise data typically encountered at LHC. Given a data point $p(1\pm \sigma)$, with $p$ being the central value and $\sigma$ its uncorrelated relative uncertainty, corresponding to some values of $q_T$ and $Q$, we include it in the fit only if
\begin{eqnarray}\label{DY-data:cuts}
\delta\equiv \frac{\langle q_T\rangle}{\langle Q\rangle } <0.1,\qquad \text{or}\qquad \delta<0.25\quad \text{if}\quad \delta^2<\sigma.
\end{eqnarray}
In other words, if the (uncorrelated) experimental uncertainty of a given data point is smaller than the theoretical uncertainty associated to the expected size of the power corrections, we drop it from the fit. 

The resulting data set in tab.~\ref{tab:data} contains 507 data points, and spans a wide range in mass, from $Q=4$~GeV to $Q=150$~GeV, and in $x$, from $x\sim0.5\cdot10^{-4}$ to $x\sim 1$.  The data roughly split into low-energy and high-energy points. The low-energy subset contains data from fixed target experiments E288 \cite{Ito:1980ev}, E605 \cite{Moreno:1990sf} and E772 \cite{McGaughey:1994dx}, and PHENIX \cite{Aidala:2018ajl}. The high-energy subset contains data from the neutral current DY (Z/$\gamma$-boson) measured at the Tevatron  \cite{Affolder:1999jh,Aaltonen:2012fi,Abbott:1999wk,Abazov:2007ac,Abazov:2010kn} and 
LHC \cite{Aad:2014xaa,Aad:2015auj,Chatrchyan:2011wt,Khachatryan:2016nbe,CMS:2019raw,Aaij:2015gna,Aaij:2015zlq,Aaij:2016mgv}. The subsets have similar number of points but those in the high-energy subset are one order of magnitude more precise. The data that we are using are very similar to those included in previous fits \cite{Bertone:2019nxa,Scimemi:2019cmh,Bacchetta:2019sam} but we 
consider also the recent CMS 
Z-boson production data at 13 TeV \footnote{We learned about these data when the main fit had already been made. Therefore, the CMS Z-boson production data at 13 TeV 
are not included in the fit but only in the comparisons.} \cite{CMS:2019raw}.

The treatment of particular aspects of the measurements such as bin-integration, nuclear modifications and normalization conditions  
are as in ref.  \cite{Scimemi:2019cmh} and we refer to it for extra details. Notice that we use the absolute values of the cross-section, whenever available, and 
perform the full bin-integration to accurately incorporate the phase-space effects. 
The only modification of the data treatment in comparison to  \cite{Scimemi:2019cmh} is the splitting of the low-energy measurements (namely E288 at $E_{\text{beam}}=300$ and $400$ GeV, E605 and E772) into two independent subsets below and above the $\Upsilon$-resonance. This allows us to treat the normalization error independently for these energies. The reason for doing  this is a possible inconsistency  between these energy regions inside the measurements, which leads to tensions. 

\begin{table}[t]
\begin{center}
\small
\begin{tabular}{|c||c|c|c|c|c|c|}
\hline
Experiment & ref. 
&$\sqrt{s}$ [GeV]& $Q$ [GeV] & $y$/$x_F$ & \specialcellcenter{$N_{\rm pt}$after \\TMD cuts}
& \specialcellcenter{$N_{\rm pt}$\\in red.set}
\\\hline\hline 
E288 (200) & \cite{Ito:1980ev} 
& 19.4 & \specialcellcenter{4 - 9 in\\ 1~GeV bins$^*$} & $0.1<x_F<0.7$  & 43 & 39
\\\hline
E288 (300) & \cite{Ito:1980ev} 
& 23.8 & \specialcellcenter{4 - 12 in \\ 1~GeV bins$^*$} & $-0.09<x_F<0.51$ & 53 & 53
\\\hline
E288 (400) & \cite{Ito:1980ev} 
& 27.4 & \specialcellcenter{5 - 14 in \\ 1~GeV bins$^*$} & $-0.27<x_F<0.33$ & 76 & 76
\\\hline\hline
E605 & \cite{Moreno:1990sf} 
& 38.8 & \specialcellcenter{7 - 18 in \\ 5 bins$^*$} & $-0.1<x_F<0.2$ & 53 & 53
\\\hline\hline
E772 & \cite{McGaughey:1994dx} 
& 38.8 & \specialcellcenter{5 - 15 in \\ 8 bins$^*$} & $0.1<x_F<0.3$ & 35 & 24
\\\hline\hline
PHENIX & \cite{Aidala:2018ajl} 
& 200 & 4.8 - 8.2 & $1.2<y<2.2$ & 3 & 2
\\\hline\hline
CDF (run1) & \cite{Affolder:1999jh} 
& 1800 & 66 - 116 & inclusive & 33 & 0
\\\hline
CDF (run2) & \cite{Aaltonen:2012fi} 
& 1960 & 66 - 116 & inclusive & 39 & 15
\\\hline\hline
D0 (run1) & \cite{Abbott:1999wk} 
& 1800 & 75 - 105 & inclusive & 16 & 0
\\\hline
D0 (run2) & \cite{Abazov:2007ac} 
& 1960 & 70 - 110 & inclusive & 8 & 0
\\\hline
D0 (run2)$_\mu$ & \cite{Abazov:2010kn} 
& 1960 & 65 - 115 & inclusive & 3 & 3
\\\hline\hline
ATLAS (7TeV) & \cite{Aad:2014xaa} 
& 7000 & 66 - 116 & \specialcellcenter{$|y|<1$\\ $1<|y|<2$ \\ $2<|y|<2.4$} &  15 & 0
\\\hline
ATLAS (8TeV) & \cite{Aad:2015auj} 
& 8000 & 66 - 116 & \specialcellcenter{$|y|<2.4$\\ in 6 bins} & 30 & 30
\\\hline
ATLAS (8TeV) & \cite{Aad:2015auj} 
& 8000 & 46 - 66 & $|y|<2.4$ &  3 & 3
\\\hline
ATLAS (8TeV) & \cite{Aad:2015auj} 
& 8000 & 116 - 150 & $|y|<2.4$ & 7 & 0
\\\hline\hline
CMS (7TeV) & \cite{Chatrchyan:2011wt} 
& 7000 & 60 - 120 & $|y|<2.1$ & 8 & 0
\\\hline
CMS (8TeV) & \cite{Khachatryan:2016nbe} 
& 8000 & 60 - 120 & $|y|<2.1$ & 8 & 0
\\\hline
CMS (13TeV) & \cite{CMS:2019raw} 
& 13000 & 76 - 106 & \specialcellcenter{$|y|<2.4$\\ in 6 bins} & 50 & 0$^{**}$
\\\hline\hline
LHCb (7TeV) & \cite{Aaij:2015gna} 
& 7000 & 60 - 120 & $2<y<4.5$ & 8 & 4
\\\hline
LHCb (8TeV) & \cite{Aaij:2015zlq} 
& 8000 & 60 - 120 & $2<y<4.5$ & 7 & 7
\\\hline
LHCb (13TeV) & \cite{Aaij:2016mgv} 
& 13000 & 60 - 120 & $2<y<4.5$ & 9 & 0
\\\hline\hline
Total & & 
& & & 507 & 309
\\\hline
\end{tabular}
\par
*Bins with $9\lesssim Q \lesssim 11$ are omitted due to the $\Upsilon$ resonance.
\caption{\label{tab:data}
Summary table for the data included in the fit. For each data set we report: reference publication, centre-of-mass energy, coverage in $Q$ and $y$ or $x_F$, possible cuts on the fiducial region, and number of data points that survive the cut of eq.~(\ref{DY-data:cuts}).}
\end{center}
\end{table}

\subsection{Reduced data set}

Given the data set defined above, we observe that the sensitivity to NP parameters of the TMD distributions varies strongly within the data set. For instance, 
data points   with $q_T\sim$ 15-25 GeV   have little power in constraining  NP parameters, irrespective of their precision, because 
they are deeply inside the resummation region, where the Hankel integral is dominated by the $b \ll 1$GeV$^{-1}$ contribution, and determined by PDF values only. The inclusion of these points in the fit is not harmful but rather time consuming. To speed up the fitting procedure, we have considered a reduced set of data points.

To identify the relevant points for the TMD extraction, we tested the sensitivity of the theory prediction to the variation of NP parameters, and included only the points that are sensitive to the variation. Given a NP parameter $p$, we compute the cross-section for several values of $p$ distributed in the range $p\pm\delta p$, and compute the sensitivity coefficient between $p$ and the cross-section, by the formula
\begin{eqnarray}
s_{\sigma,p}=\frac{\langle \sigma \cdot p \rangle-\langle \sigma\rangle \langle p\rangle}{\Delta \sigma \, \delta p},
\end{eqnarray}
where $\Delta \sigma$ is the uncorrelated experimental uncertainty of the point. The sensitivity coefficient indicates how strongly a prediction for a data point depends on the parameter $p$. E.g., $s_{\sigma,p}=1$ means that the variation of $p$ by $\delta p$ gives rise to a variation in the theory prediction equal to the experimental uncertainty.

We computed the sensitivity coefficient for each point in the data set and for each parameter of our NP ansatz. The values of $\delta p$ are taken using 
the Hesse estimation for its uncertainty multiplied by a factor 5. Points with $s>0.4$ for at least one parameter were included in the reduced set. This test was run for each 
 of the PDF sets that we studied, and the union of data was taken. This provided us with a very conservative selection of impacting data.

Afterwards, we excluded the sets with too few points, such as LHCb (13 TeV) (1 surviving point) and CDF (run1) (5 points remaining out of 33). Additionally, we excluded the 
ATLAS (7 TeV) measurement since it has shown an anomalous behaviour and provides significant tension with other similar measurements\footnote{The same observation was also made by the PDF fitting collaborations NNPDF \cite{Ball:2017nwa} and CT-TEA \cite{Hou:2019efy}, which exclude these data from the pool.}.

The reduced set contains 309 data points. Most excluded points are from the region characterized by high energy and high $q_T$. The low-energy subset, in contrast, is very sensitive to the NP input and lost only 15 points due to their large error bands. We have checked, by evaluating several random cases, that the minimum of the $\chi^2$ of the reduced set almost coincides (up to 2 digits) with the minimum of  $\chi^2$ of the complete set. 

\section{Statistical framework}
\label{sec:statistical}

The central point of the present study is the impact of the PDF uncertainty on the extraction of TMD distributions. Therefore, we distinguish two sources of uncertainties: $i)$ the experimental uncertainties and $ii)$ the uncertainty of the collinear PDFs. These have different nature, and could not be easily combined together. In fact, the best approach to account for the PDF uncertainty would be a simultaneous global fit of collinear PDF and TMDPDF, which is challenging. In the present work, we use the strategy of combining PDF and experimental uncertainty based on Bayesian statistics, described in this section. 

\subsection{Definition of the $\chi^2$-function}

Our main tool is the $\chi^2$-test function, which estimates the goodness of a theory prediction. It is defined as
\begin{equation}\label{eq:chi2cov}
\chi^2=\sum_{i,j=1}^{n}\(m_i-t_i\right)V_{ij}^{-1}\(m_j-t_j\)\, ,
\end{equation}
where $i$ and $j$ enumerate points of the data set. For the $i$th point, $t_i$ is the corresponding theoretical prediction, $m_i$ is the experimental value, and all the information about the uncertainties for individual data points and the relation between them is given by the covariance matrix $V_{ij}$. The definition of $V_{ij}$ is crucial for an adequate analysis of the data and interpretation of the results. Its construction distinguishes two types of uncertainties: uncorrelated and correlated. In general, the $i$th point is given as 
\begin{equation}
m_i\pm \sigma_{i,\text{stat}} \pm \sigma_{i,\text{unc}} \pm \sigma_{i,\text{corr}}^{(1)}\pm\dots \pm \sigma_{i,\text{corr}}^{(k)} \, .
\end{equation}
Here $\sigma_{i,\text{stat}}$ and $\sigma_{i,\text{unc}}$ are the uncorrelated  statistical and systematic uncertainties, respectively, that estimate the degree of knowledge of the $i$th data point regardless of all other data in the set. The correlated uncertainties $\sigma_{i,\text{corr}}^{(k)}$ (with $k=1,...,n$) estimate the relation between the statistical fluctuations of the $i$th point and all others in the set. The covariance matrix $V_{ij}$ can be written as \cite{Ball:2008by,Ball:2012wy}
\begin{equation}\label{eq:covmat}
V_{ij}=\(\sigma_{i,\text{stat}}^2 +\sigma_{i,\text{unc}}^2\)\delta_{ij} + \sum_{l=1}^{k}\sigma_{i,\text{corr}}^{(l)}\sigma_{j,\text{corr}}^{(l)}, .
\end{equation}
Using this expression in the computation of eq.~(\ref{eq:chi2cov}) allows us to obtain a  reliable estimation of data-to-theory agreement, while taking into account 
the nature of experimental uncertainties.

The minimum of the $\chi^2$-test function determines the preferred values of the NP parameters $\lambda$, which in turn characterize the preferred shape of TMD distributions. The fit procedure consists in the minimization of $\chi^2$, which is performed with the iMinuit package \cite{iminuit}.

In fitting a multivariate function one usually finds several sets of parameters giving numerous local minima of the function. Some of these give mathematically correct but physically unsound results, and therefore a strategy should be implemented to discard them. The MINUIT algorithm can handle the problem of choosing the correct minimum, unless the minima are significantly separated in the parameter space and/or the step (the amount by which the parameters move in one iteration) is too small. In our case, with the $\chi^2$-function depending on 11 parameters, we have faced the problem of the minimisation algorithm selecting a definitely unphysical minimum. Each occurrence is characterized by values of parameters that (almost) suppress a contribution of some flavor. 

In order to disregard such contributions we introduced penalty terms into the minimisation function. The penalty function is defined as
\begin{eqnarray}
P(\lambda)=10^{-2}\(
r(\lambda^u_1,\lambda_1^d)
+r(\lambda^u_2,\lambda_2^d)
+r(\lambda^{\bar u}_1,\lambda_1^{\bar d})
+r(\lambda^{\bar u}_2,\lambda_2^{\bar d})\),
\end{eqnarray}
where
\begin{eqnarray}
r(a,b)=\left\{
\begin{array}{ll}
x-1, & x>1\\
0,  & \text{otherwise}, 
\end{array}
\right.\qquad \text{with}\quad x=\max\(\frac{a}{b},\frac{b}{a}\).
\end{eqnarray}
This function is non-zero only if the parameters of the $u$/$d$ and $\bar u$/$\bar d$ distributions differ from each other by an order of magnitude. So  the 
minimisation was performed for the function
\begin{eqnarray}
X(\text{data},\text{PDF};\lambda)=\chi^2+P(\lambda).
\end{eqnarray}

\subsection{Input PDFs and their uncertainties}
\label{sec:PDF-replicas}

\begin{figure}[t]
\centering
\includegraphics[width=0.89\textwidth]{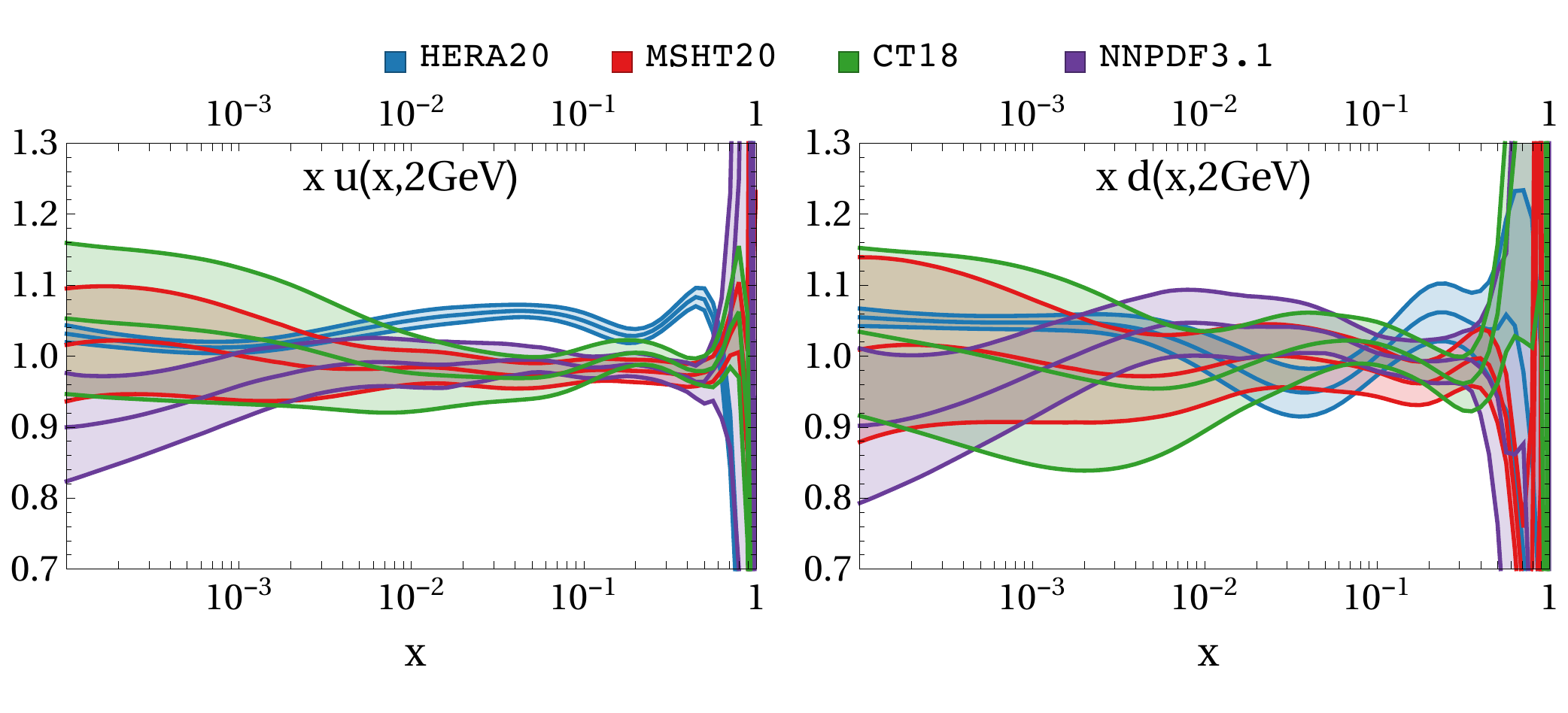}
\caption{\label{fig:compare-PDF}
Comparison of the uncertainty bands of  PDFs extracted by different groups for $u$ and $d$ quarks. The PDFs are weighted by an average of the central values. }
\end{figure}

Four PDF sets were used in our study, the bold font denoting the shorthand name used to identify them in the rest of this article: 
\begin{itemize}
\item \textbf{HERA20}. The NNLO extraction by the H1 and ZEUS collaborations presented in ref.~\cite{Abramowicz:2015mha} with Hesse-like error band. The LHAPDF entry is \texttt{HERAPDF20_NNLO_VAR} with id = 61230.
\item \textbf{NNPDF31}. The NNLO extraction by the NNPDF collaboration presented in ref.~\cite{Ball:2017nwa} with 1000 replicas at $\alpha_s(M_Z)=0.118$. The LHAPDF entry is \texttt{NNPDF31\_nnlo\_0118\_1000} with id = 309000.
\item \textbf{CT18}. The NNLO extraction by the CTEQ collaboration presented in ref.~\cite{Hou:2019efy} with Hesse error band. The LHAPDF entry is \texttt{CT18NNLO} with id = 14000.
\item \textbf{MSHT20}. The NNLO extraction by the MSHT collaboration presented in ref.~\cite{Bailey:2020ooq} with Hesse error band. The LHAPDF entry is \texttt{MSHT20nnlo\_as118} with id = 27400.
\end{itemize}
The comparison of these PDF sets for $u$- and $d$-quarks at scale $\mu = 2 $ GeV is presented in fig.~\ref{fig:compare-PDF}.

In what follows the analyses are done using Bayesian statistics, which requires  representing the PDFs as Monte-Carlo (MC) ensembles. As the \textbf{NNPDF31} set is already given in this form, no further pre-processing is required. The other three distributions have a Hessian definition of uncertainty bands. The corresponding MC ensembles are generated using the prescription given in ref.~\cite{Hou:2016sho}. Namely, for a distribution $f(x)$ with 68\% C.I. for each eigenvector given by 
$f^\pm_i$ ($i=1, \dots , D$, with $f^+_i$ and $f^-_i$ the upper and lower bounds, respectively), a MC replica is generated by
\begin{eqnarray}
f^{(k)}(x)=\sum_{i=1}^D \(\frac{f^+_i(x)-f^-_i(x)}{2}R_i^{(k)}+
\frac{f^+_i(x)+f^-_i(x)-2f(x)}{2}\(R_i^{(k)}\)^2\),
\end{eqnarray}
where $R^{(k)}_i$ is a univariate random number. Using this method we generated 1000 MC replicas for \textbf{HERA20}, \textbf{CT18} and \textbf{MSHT20}, which were used in the analyses. We have checked that the uncertainty bands obtained from the generated MC replicas are almost identical to the original Hesse uncertainty bands. The comparison of uncertainty bands for different PDF sets is also 
shown in fig.\ref{fig:compare-PDF}.

\subsection{Determination of uncertainties \& presentation of the result}
\label{sec:EXP+PDF}

Within Bayesian statistics, the propagation of uncertainties to the free parameters is made by fitting each member of the input ensemble. Our input ensemble is generated accounting for two independent sources of uncertainty:
\begin{itemize}
\item[] \textbf{EXP}:  The experimental uncertainty is accounted for by generating pseudo-data. A replica of the  pseudo-data is obtained adding Gaussian noise to the values of the data points (and scaling the uncertainties if required). The parameters of the noise are dictated by the correlated and uncorrelated experimental uncertainties. The procedure is described in ref.~\cite{Ball:2008by}. We considered such 100 replicas.
\item[] \textbf{PDF}: The uncertainty due to the collinear PDF is accounted for by using each PDF replica as input. We considered 1000 replicas generated as described in sec. \ref{sec:PDF-replicas}.
\end{itemize}
The number of pseudo-data replicas is lower because the resulting uncertainty is much smaller than the one coming from the PDF distribution, as demonstrated in the following. 

Fitting each member of the ensemble, we end up with the two-dimensional set $\vec \lambda_{ij}$, where $\vec \lambda$ is a 11-dimensional vector of fitting parameters (including $\lambda_{1,2}^f$, $\lambda_0$ and $c_0$), $i$ runs over the pseudo-data replicas, and $j$ runs over the PDF replicas. This set is distributed in accordance to the experimental and PDF uncertainties propagated through our fitting procedure. Due to the computational limitations mentioned in subsec. \ref{subsec:nonpertinputs}, calculating the full distribution with $100\times 1000=10^5$ members is unrealistic. To simplify the task, we consider two distributions. The first one, labeled {\bf EXP}, when the replicas of the data are fitted with the central PDFs. The second one, called {\bf PDF}, when the replicas of the PDFs are fitted with the central (original) data. Symbolically,
\begin{eqnarray}
\{\vec \lambda\}_{\textbf{PDF}}=\bigcup_{j=1}^{N_{\text{rep,PDF}}}\vec \lambda_{0j},
\qquad
\{\vec \lambda\}_{\textbf{EXP}}=\bigcup_{i=1}^{N_{\text{rep,data}}}\vec \lambda_{i0},
\end{eqnarray}
where $i=0$ indicates the replica with unmodified data, and $j=0$ a replica computed with the central value of the PDF.

The {\bf PDF} and {\bf EXP} distributions are treated as independent. 
In the {\bf PDF} case the distributions are also notably non-Gaussian. Therefore, we estimate the 68\%C.I. for a parameter using the bootstrap method. Namely, we compute the [16\%, 84\%] quantile for a large number of samples and take the average interval. The results are presented as $a^{\delta a_1}_{\delta a_2}$, where $a$ is the mean value of the parameter and $\delta a$'s are the distances to the 68\% C.I. boundary.

A drawback of performing the fit independently for each case is the appearance of several issues when wanting to join the two forms in a single meaningful one. The main problem is that, for some parameters, the distance between the mean values of the distributions is large. This renders impossible any naive description of a joined distribution; for example, the average value of the extracted TMDPDFs would have a large $\chi^2$, meaning that the corresponding TMDPDFs would not provide an adequate description of the data. Lesser, but not unimportant, problems are the correlations between individual points of TMDPDFs, and the PDF bias of the {\bf EXP} case. Therefore, finally we present the central value and the 68\% C.I., computed 
as described below. 

The central value of our fit is obtained as the TMDPDF computed with the central values of the PDFs and NP parameters. The latter are computed as the weighted average of the {\bf PDF} and {\bf EXP} cases, 
\begin{eqnarray}
a=w_{\textbf{PDF}}\langle \{a\}_{\textbf{PDF}}\rangle+w_{\textbf{EXP}}\langle \{a\}_{\textbf{EXP}}\rangle,\qquad w_i=\frac{\sigma^{-2}_i}{\sigma^{-2}_{\textbf{PDF}}+\sigma^{-2}_{\textbf{EXP}}},
\end{eqnarray}
where $i=\text{\bf PDF}$ or $\text{\bf EXP}$, and $\sigma_i=(\delta a_{1i}-\delta a_{2i})/2$ is the half-size of the 68\% C.I.. In this way, the central value incorporates information from both cases but retains the knowledge of which case gives a better determination of the considered parameter. The final parameters obtained by this procedure for each PDF set are shown in tab. \ref{tab:parameters}. Also, the central distribution so defined returns a reasonable value of $\chi^2$, which should be similar to the result that we would obtain if we had performed one joined fit. The joined 68\% C.I., instead, is computed from the (equal weight) sum of replicas distributions, $\{\vec \lambda\}_{\text{joined}}=\{\vec \lambda\}_{\textbf{PDF}}\cup \{\vec \lambda\}_{\textbf{EXP}}$ by the bootstrap method. 

The same procedure is also carried out for the values of cross-section presented in the next section. I.e. the values of cross-section are computed for all members of  $\{\vec \lambda\}_{\textbf{PDF}}$ and $\{\vec \lambda\}_{\textbf{EXP}}$ resulting into $\{\sigma\}_{\textbf{PDF}}$ and $\{\sigma\}_{\textbf{EXP}}$. Then the central values and the 68\% C.I. are computed using the same strategy. That is, we compute the [16\%, 84\%] quantile for a large number of samples drawn from the merge of {\bf PDF} and {\bf EXP} and take the average interval. For those parameters for which {\bf PDF} and {\bf EXP} coincide, the width of the sampled distributions will be narrow, the contribution to the total uncertainty closely following the one of {\bf EXP} (red band in fig. \ref{fig:example:high-energy}). For those parameters for which {\bf PDF} and {\bf EXP} do not significantly overlap, the width of the sampled distributions will be broader, the contribution to the total uncertainty closer, but narrower, to the one from {\bf PDF} (green band in fig. \ref{fig:example:high-energy}). Overall, the bootstrapping method results in an uncertainty band that resembles the one we would expect if performing the fit with simultaneous replicas of the data and the PDFs. This procedure (rather than computation of average values of parameters) accounts for the correlation between different values of TMDPDFs.

\section{Results \& discussion}
\label{sec:results}

\begin{figure}[t]
\centering
\includegraphics[width=0.99\textwidth]{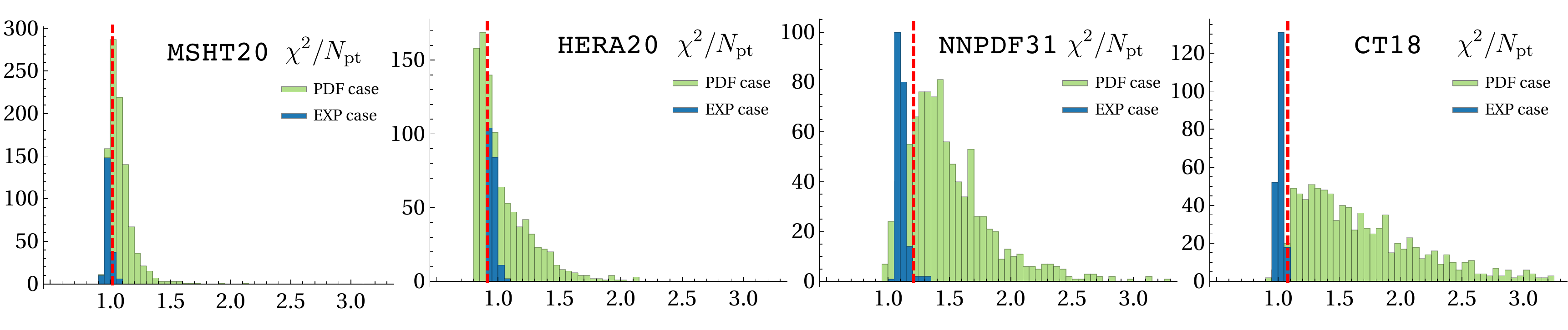}
\caption{\label{fig:chi2}
Distribution of $\chi^2$-values for the {\bf PDF} and {\bf EXP} cases. The red lines show the position of the final $\chi^2$-value.}
\end{figure}

\begin{table}[t]
\footnotesize
\begin{center}
\renewcommand{\arraystretch}{0.94}
\begin{tabular}{|l | c V{4} c V{4} c V{4} c V{4} c V{4}}
\cline{3-6}
\multicolumn{2}{c V{4}}{ } & 
\textbf{MSHT20}  &
\textbf{HERA20}  &
\textbf{NNPDF31}  &
\textbf{CT18}  \\
\hline 
Data set & $N_{pt}$ & 
$\chi^2/N_{pt}$ & 
$\chi^2/N_{pt}$ & 
$\chi^2/N_{pt}$ & 
$\chi^2/N_{pt}$  
\\\hline
CDF run2          &   15 &      0.96 & 0.68 & 0.65 & 0.82 \\
D0 run2 ($\mu$)   &    3 &      0.50 & 0.59	& 0.55 & 0.52 \\
\hline
ATLAS 8TeV 0.0<|y|<0.4      &    5 &      2.97 & 3.66 & 2.12 & 3.23\\
ATLAS 8TeV 0.4<|y|<0.8      &    5 &      2.00 & 1.53 & 4.52 & 3.21\\
ATLAS 8TeV 0.8<|y|<1.2      &    5 &      1.00 & 0.50 &	2.75 & 1.89\\
ATLAS 8TeV 1.2<|y|<1.6      &    5 &      2.25 & 1.61 & 2.49 & 2.72\\
ATLAS 8TeV 1.6<|y|<2.0      &    5 &      1.92 & 0.68 &	2.86 & 1.96\\
ATLAS 8TeV 2.0<|y|<2.4      &    5 &      1.35 & 1.14 & 1.47 & 1.06\\
ATLAS 8TeV 46<Q<66GeV      &    3 &       0.59 & 1.86 & 0.23 & 0.05\\
\hline
LHCb 7TeV         &    4 &      3.19 & 0.34 & 2.58 & 1.68\\
LHCb 8TeV         &    7 &      1.38 & 1.29 & 1.63 & 0.83\\
\hline
PHE200             &    2 &      0.32 &	0.36 & 0.43 & 0.27 \\
E228-200           &   39 &      0.44 &	0.38 & 0.51 & 0.45 \\
E228-300 $Q<9$GeV  &   43 &      0.77 & 0.56 & 0.89 & 0.55 \\
E228-300 $Q>11$GeV &   10 &      0.29 & 0.37 & 0.45 & 0.44 \\
E228-400 $Q<9$GeV  &   34 &      2.19 & 1.15 & 1.49 & 1.34 \\
E228-400 $Q>11$GeV &   42 &      0.25 & 0.61 & 0.44 & 0.40 \\
E772               &   24 &      1.58 & 1.92 & 2.51 & 1.56 \\
E605     $Q<9$GeV  &   21 &      0.52 &	0.47 & 0.47 & 0.61 \\
E605     $Q>11$GeV &   32 &      0.47 & 0.73 & 1.34 & 0.52 \\
\hline\hline
\textbf{Total} & \textbf{309} & \textbf{0.97} & \textbf{0.85} & \textbf{1.17} & \textbf{0.87}\\
\hline
\end{tabular}
\caption{\label{tab:final-redu-XSV21} Distribution of the values of $\chi^2$ for the central replica  over the reduced data set in fits with different PDF inputs. }
\end{center}
\end{table}

\begin{table}
\footnotesize
\begin{center}
\renewcommand{\arraystretch}{0.94}
\begin{tabular}{|l | c V{4} c V{4} c V{4} c V{4} c V{4}}
\cline{3-6}
\multicolumn{2}{c V{4}}{ } & 
\textbf{MSHT20}  &
\textbf{HERA20}  &
\textbf{NNPDF31}  &
\textbf{CT18}  \\
\hline 
Data set & $N_{pt}$ & 
$\chi^2/N_{pt}$ & 
$\chi^2/N_{pt}$ & 
$\chi^2/N_{pt}$ & 
$\chi^2/N_{pt}$  
\\\hline
CDF run1          &   33 &      0.78 & 0.61	& 0.72 	& 0.75 \\
CDF run2          &   39 &      1.70 & 1.42	& 1.68 	& 1.79 \\
D0 run1           &   16 &      0.71 & 0.81 & 0.79  & 0.79 \\
D0 run2           &    8 &      1.95 & 1.39	& 1.92  & 2.00 \\
D0 run2 ($\mu$)   &    3 &      0.50 & 0.59	& 0.55  & 0.52 \\
\hline
ATLAS 7TeV 0.0<|y|<1.0      &    5 &      4.06 & 1.94 & 2.12 & 4.21\\
ATLAS 7TeV 1.0<|y|<2.0      &    5 &      7.78 & 4.83 & 4.52 & 6.12\\
ATLAS 7TeV 2.0<|y|<2.4      &    5 &      2.57 & 2.18 & 3.65 & 2.39\\
ATLAS 8TeV 0.0<|y|<0.4      &    5 &      2.98 & 3.66 & 2.12 & 3.23\\
ATLAS 8TeV 0.4<|y|<0.8      &    5 &      2.00 & 1.53 & 4.52 & 3.21\\
ATLAS 8TeV 0.8<|y|<1.2      &    5 &      1.00 & 0.50 & 2.75 & 1.89\\
ATLAS 8TeV 1.2<|y|<1.6      &    5 &      2.25 & 1.61 & 2.49 & 2.72\\
ATLAS 8TeV 1.6<|y|<2.0      &    5 &      1.92 & 1.68 &	2.86 & 1.96\\
ATLAS 8TeV 2.0<|y|<2.4      &    5 &      1.35 & 1.14 & 1.47 & 1.06\\
ATLAS 8TeV 46<Q<66GeV      &    3 &       0.59 & 1.86 & 0.23 & 0.05\\
ATLAS 8TeV 116<Q<150GeV    &    7 &       0.61 & 1.03 & 0.85 & 0.70\\
\hline
CMS 7TeV          &    8 &      1.22 & 1.19 & 1.30 & 1.25\\
CMS 8TeV          &    8 &      0.78 & 0.77 & 0.75 & 0.78\\
CMS 13TeV 0.0<|y|<0.4 &    8 &       3.52& 1.93 & 2.13& 3.73\\
CMS 13TeV 0.4<|y|<0.8 &    8 &       1.06& 0.53 & 0.71& 1.65\\
CMS 13TeV 0.8<|y|<1.2 &    10 &      0.48& 0.14 & 0.33& 0.88\\
CMS 13TeV 1.2<|y|<1.6 &    11 &      0.62& 0.33 & 0.47& 0.86\\
CMS 13TeV 1.6<|y|<2.4 &    13 &      0.46& 0.32 & 0.39& 0.57\\
\hline
LHCb 7TeV         &    8 &      1.79 & 1.00 & 1.62 & 1.16\\
LHCb 8TeV         &    7 &      1.38 & 1.29 & 1.63 & 0.83\\
LHCb 13TeV        &    9 &      1.28 & 0.84 & 1.07 & 0.93\\
\hline
PHE200             &    3 &      0.29 &	0.42 & 0.38 & 0.29 \\
E228-200           &   43 &      0.43 & 0.36 & 0.57 & 0.43 \\
E228-300 $Q<9$GeV  &   43 &      0.77 & 0.56 & 0.89 & 0.55 \\
E228-300 $Q>11$GeV &   10 &      0.29 & 0.37 & 0.45 & 0.44 \\
E228-400 $Q<9$GeV  &   34 &      2.19 & 1.15 & 1.49 & 1.34 \\
E228-400 $Q>11$GeV &   42 &      0.25 & 0.61 & 0.44 & 0.40 \\
E772               &   35 &      1.14 & 1.37 & 1.79 & 1.11 \\
E605     $Q<9$GeV  &   21 &      0.52 & 0.47 & 0.47 & 0.61 \\
E605     $Q>11$GeV &   32 &      0.47 & 0.73 & 1.34 & 0.52\\
\hline\hline
\textbf{Total} & \textbf{507} & \textbf{1.12} & \textbf{0.91} & \textbf{1.21} & \textbf{1.08} \\
\hline
\end{tabular}
\caption{\label{tab:final-XSV21} Distribution of the values of $\chi^2$ for the central replica over the TMD data set in fits with different PDF input.}
\end{center}
\end{table}

\subsection{Agreement between data and theory}

The individual values of $\chi^2$ for each experiment obtained for the reduced and complete data sets are given in tabs.~\ref{tab:final-redu-XSV21}-\ref{tab:final-XSV21}, respectively. 
The tables also report the total $\chi^2$ of the fits and they show that
an overall reasonable description of the data is achieved with all PDF sets. Note that the ATLAS 7 TeV data have a generally higher $\chi^2$ similarly to what is found also in other TMD fits \cite{Scimemi:2019cmh,Bertone:2019nxa,Bacchetta:2019sam,Bacchetta:2022awv}. 

The $\chi^2$ distribution among the {\bf PDF} and {\bf EXP} replicas is shown in fig.~\ref{fig:chi2}. In general they are consistent with each other and their spread is highly reduced with respect to previous fits that use a flavor independent profile. More details on  this are discussed in the  appendix.
This confirms the relevance of taking into account the  flavor dependence of the NP TMD distributions  $f_{\rm NP}^f$. For all cases the {\bf PDF} replicas provide a much larger dispersion of the $\chi^2$ values than the {\bf EXP} ones. The shapes of $\chi^2$-distributions are visibly different for different collinear PDFs, despite the resulting uncertainties in cross-section and TMDPDFs being similar. This may be due to the complexity of the corresponding parameterizations. In particular, the NNPDF and CT18 cases have more disturbed shapes of individual replicas (in comparison to HERA20 and MSHT20 cases) and consequently a larger spread of the $\chi^2$ distribution.

\begin{figure}
\centering
\includegraphics[width=0.49\textwidth]{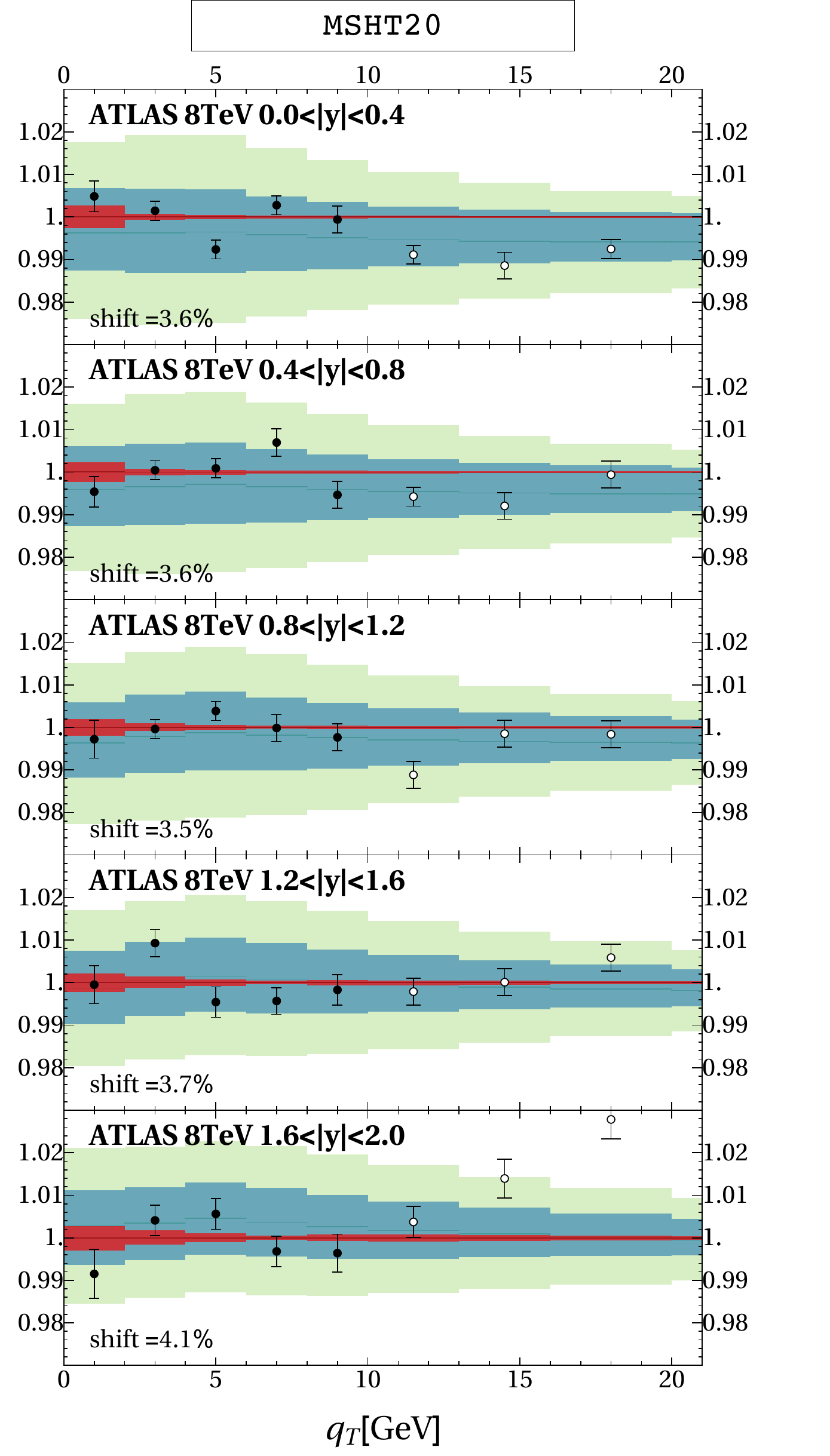}~
\includegraphics[width=0.49\textwidth]{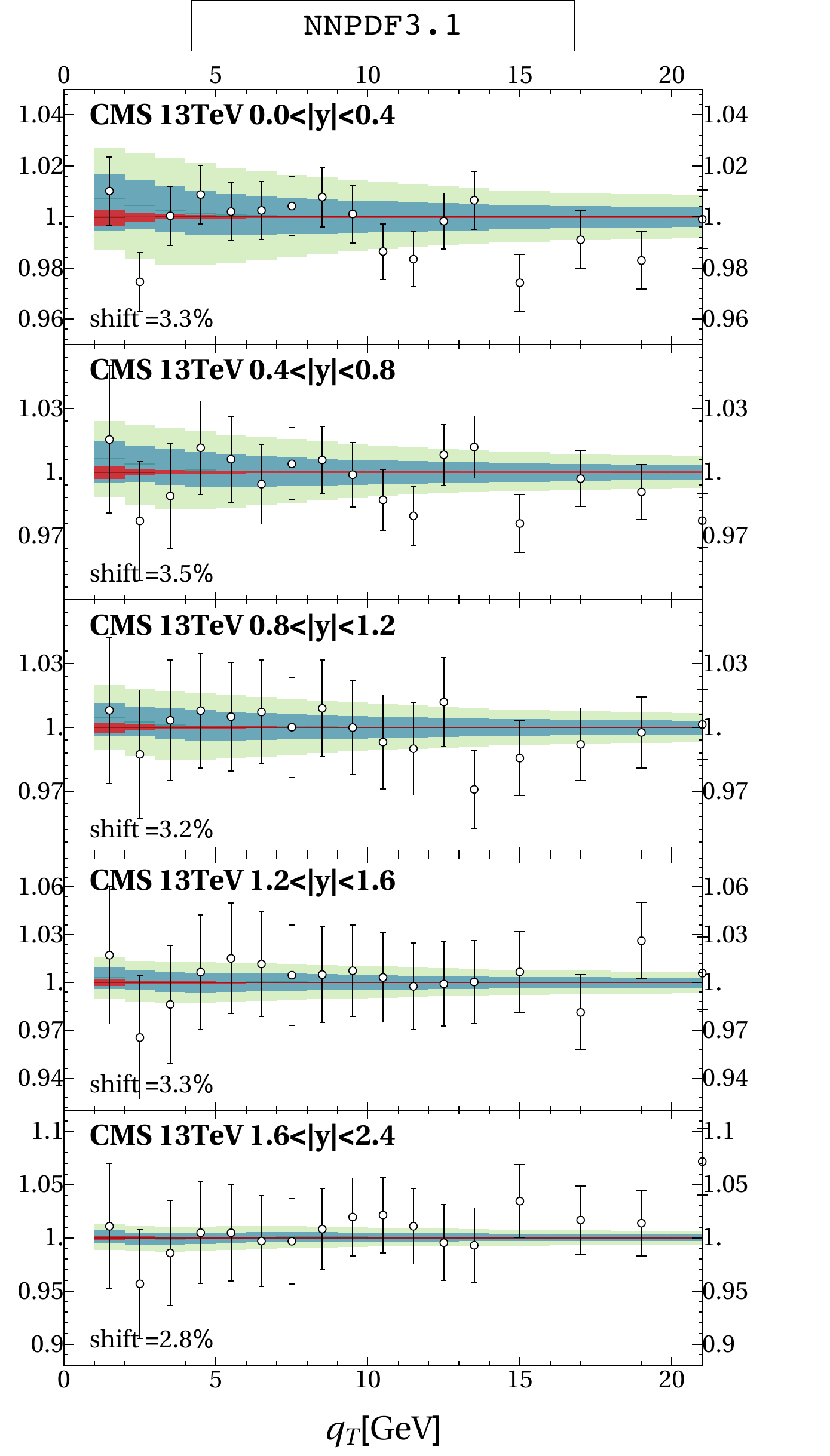}
\caption{\label{fig:example:high-energy}
Example of the data description at high energy. Left panel: the ratio $d\sigma_{\text{experiment}}/d\sigma_{\text{theory}}$ for Z-boson production at 8 TeV measured by the ATLAS experiment with MSHT20. Right panel: the ratio $d\sigma_{\text{experiment}}/d\sigma_{\text{theory}}$ for Z-boson production at 13 TeV at the CMS experiment with NNPDF3.1. The red band is the \textbf{EXP}-uncertainty. The light-green band is the \textbf{PDF}-uncertainty. The blue band is the combined uncertainty. Only the filled bullets are included into the fit.
}
\end{figure}

\begin{figure}
\centering
\includegraphics[width=0.49\textwidth]{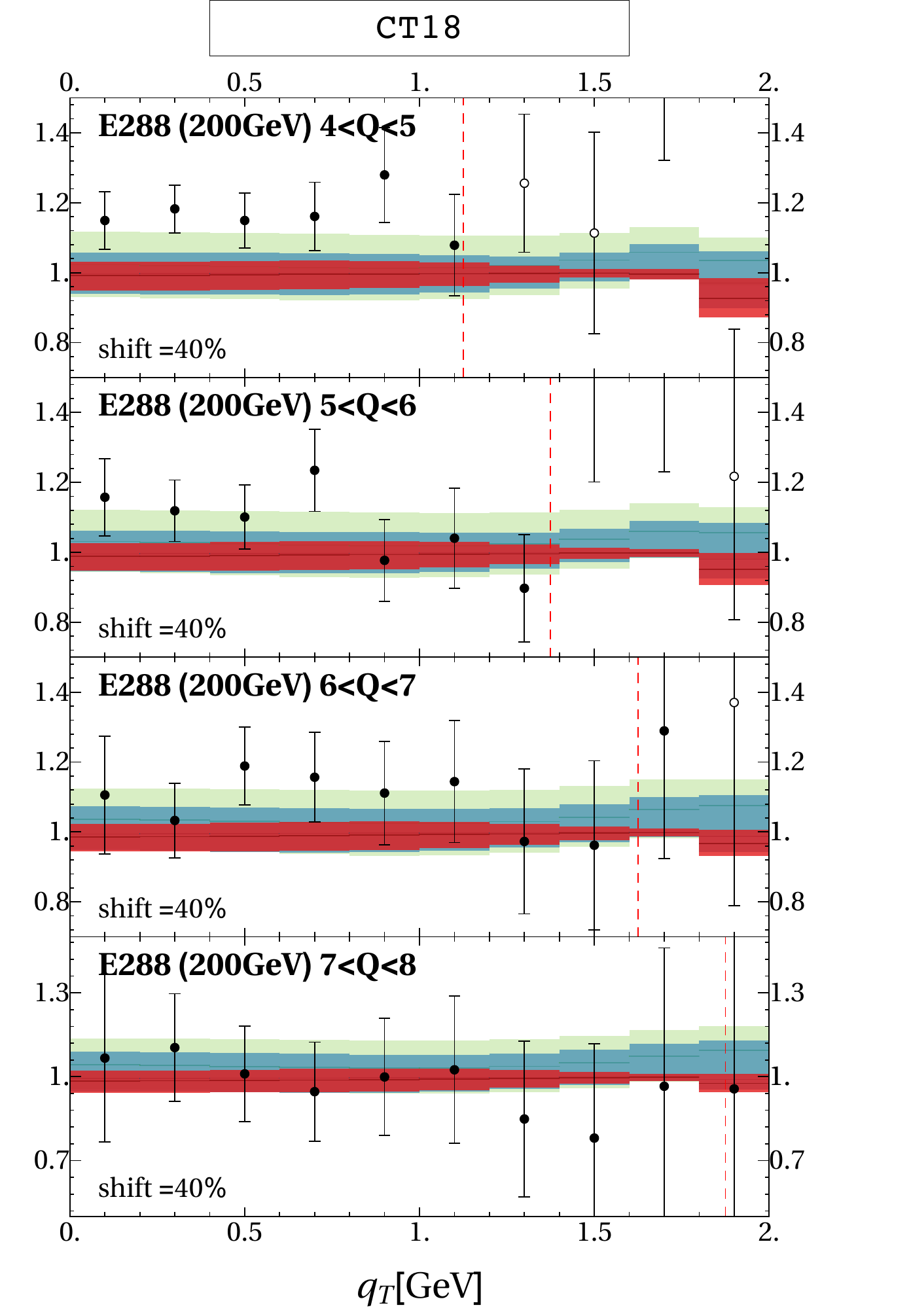}~
\includegraphics[width=0.49\textwidth]{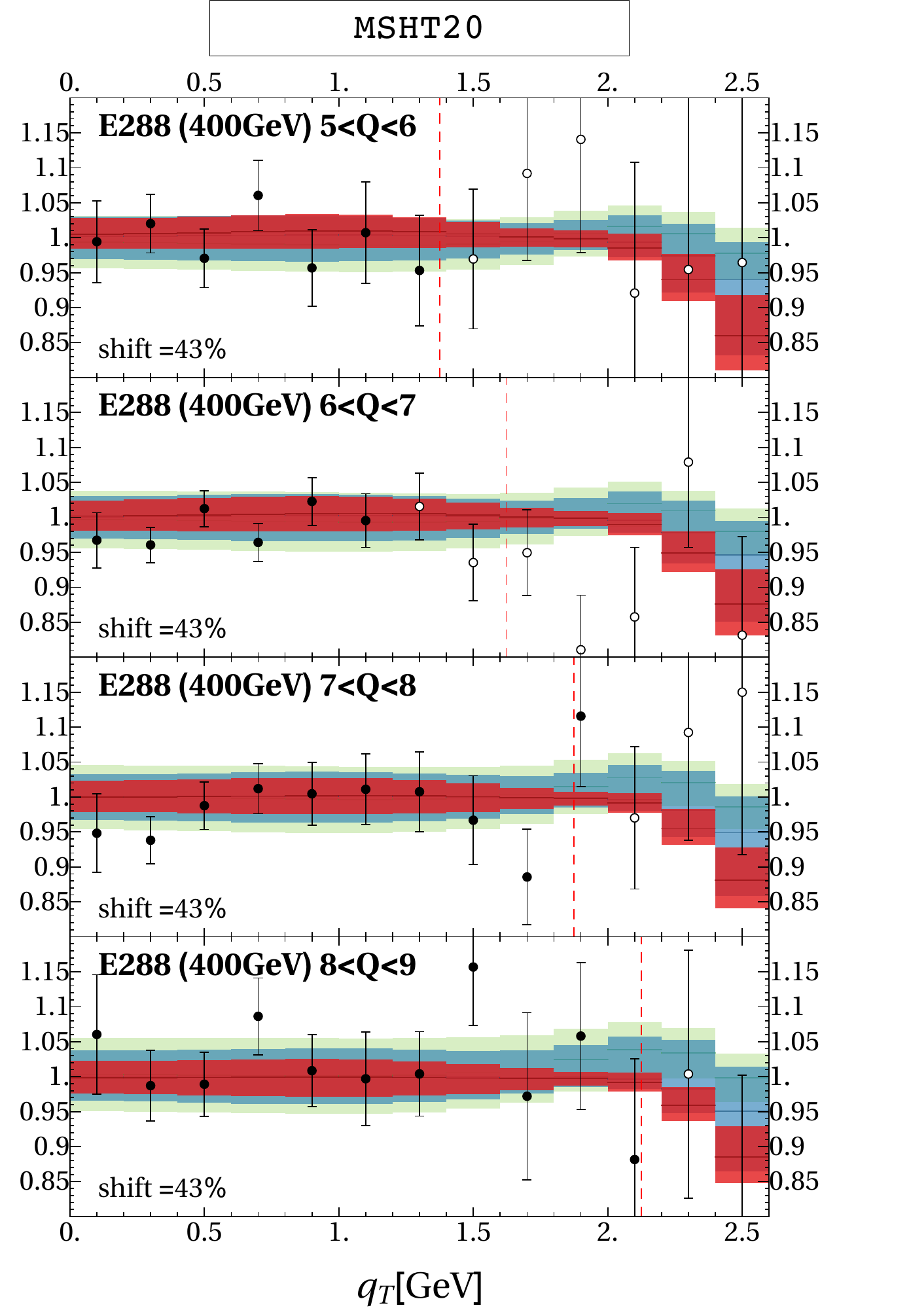}
\caption{\label{fig:example:low-energy}
Example of the data description at low energy. Left panel: ratio $d\sigma_{\text{experiment}}/d\sigma_{\text{theory}}$ for the DY process at E288 experiment with 200 GeV beam-energy with CT18. Right panel: ratio $d\sigma_{\text{experiment}}/d\sigma_{\text{theory}}$ for the DY process at E288 experiment with 400 GeV beam-energy with MSHT20. Red band is the \textbf{EXP}-uncertainty. Light-green band is the \textbf{PDF}-uncertainty. The blue band is the combined uncertainty. The filled bullets are included into the fit. The dashed red vertical lines illustrate  the cut  $q_T < 0.25Q$ discussed at the beginning of sec.~\ref{sec:data}.
}
\end{figure}

The quality achieved in describing the data within the present fit is illustrated in figs.~\ref{fig:example:high-energy}-\ref{fig:example:low-energy}, for different PDF cases. Given the variety of experiments included in the fit and the number of PDF sets used, we present here only a fraction of all obtained results. We refer the interested reader to the supplementary material for the complete collection of plots. The left panel of fig.~\ref{fig:example:high-energy} illustrates the Z-boson production measured by ATLAS at $\sqrt{s}=8$ TeV~\cite{Aad:2015auj}, the most precise data set in our analysis. On the right panel of fig.~\ref{fig:example:high-energy}, we show the comparison of the theory prediction to data for the Z-boson production measured by CMS at $\sqrt{s}=13$ TeV~\cite{CMS:2019raw}. Notice that these data were not included in the fit. As an example of the lowest energy measurements considered in the analysis, we present in fig. \ref{fig:example:low-energy} the DY-process measured by the E288 experiment~\cite{Ito:1980ev}. In all cases, the \textbf{PDF}-uncertainty is larger than the \textbf{EXP}-uncertainty. This is especially pronounced for the high-energy measurements, for which the experimental uncertainty is small, and thus the uncertainty band is dominated by the \textbf{PDF} error. For $q_T\gtrsim 10$ GeV, the \textbf{EXP}-uncertainty becomes negligible.  This is the resummation regime, insensitive to NP TMD effects. In a few cases (central rapidity Z-boson production with CT18, and the lowest energy bins for DY process with NNPDF3.1), the \textbf{PDF} and \textbf{EXP} bands do not overlap, illustrating an unrevealed tension in the fitting procedure.

\begin{figure}[t]
\centering
\includegraphics[width=0.48\textwidth]{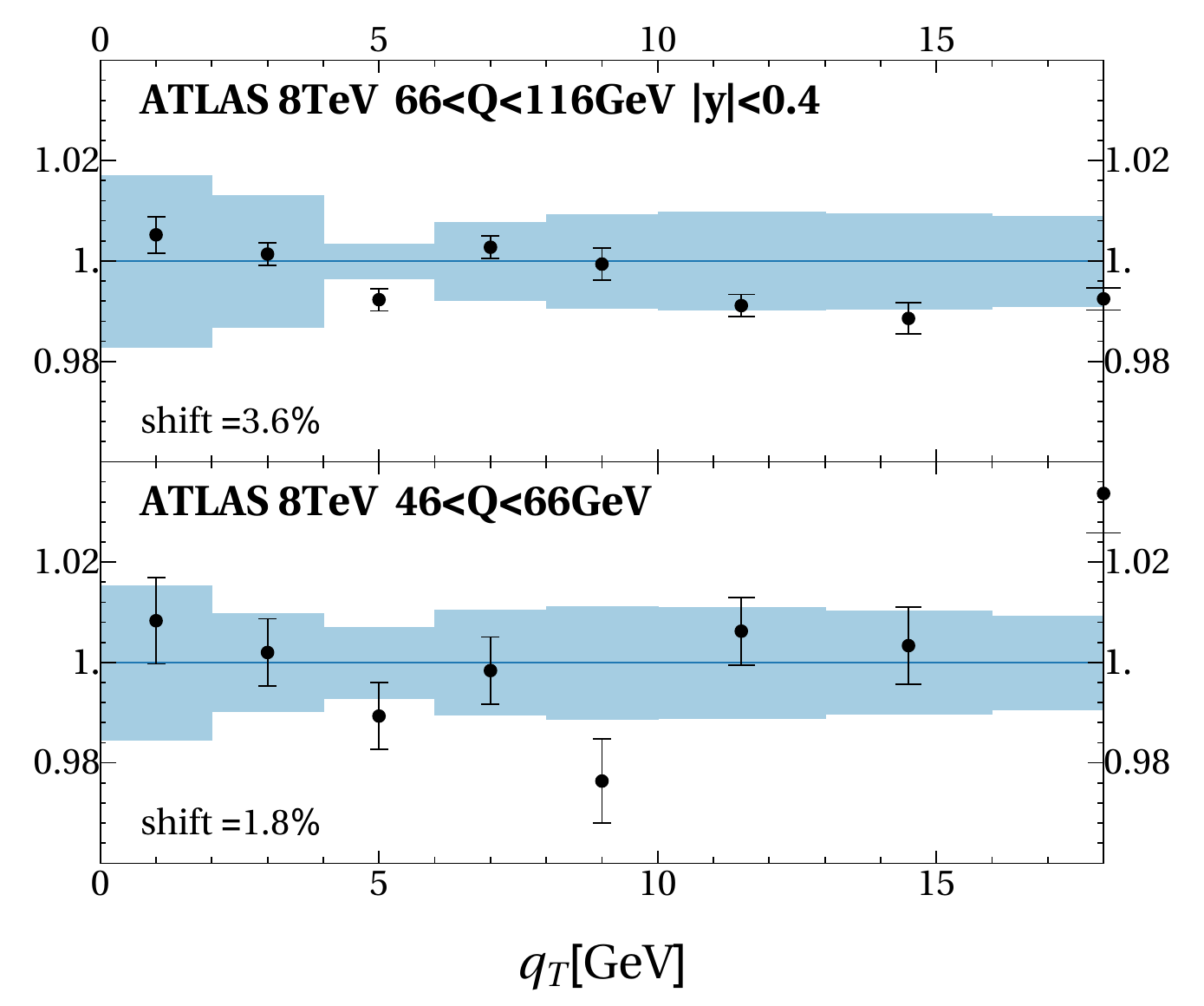}
\includegraphics[width=0.48\textwidth]{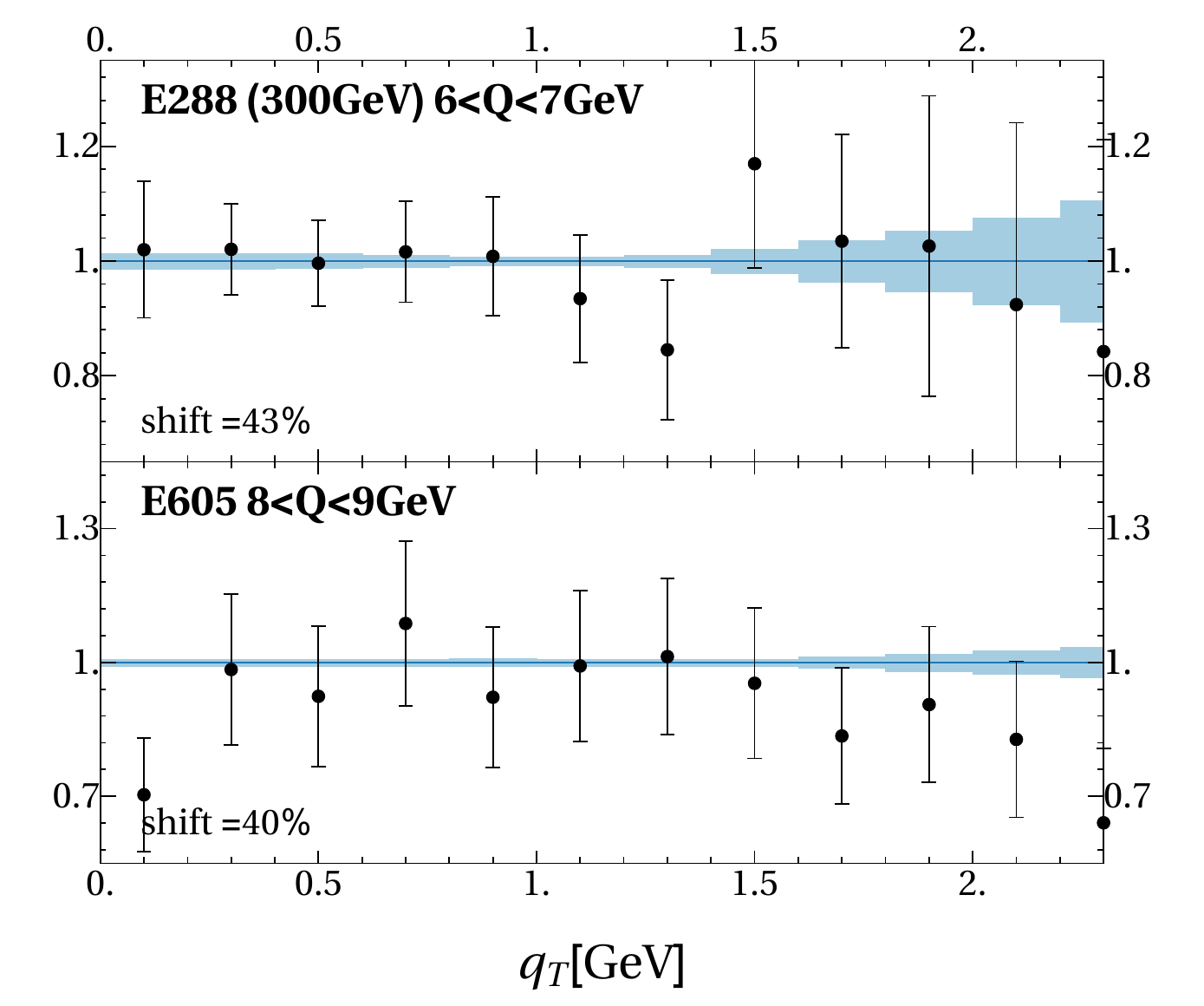}
\caption{\label{fig:scale-vary} Scale variation band in comparison to typical data at high (left panel) and low (right panel) energies. The scale variation band is defined as the maximum symmetrized deviation from varying all scale parameters by a factor $c\in[0.5,2]$.
}
\end{figure}

To better appreciate the role  of the PDF and EXP 
uncertainty bands in  figs.~\ref{fig:example:high-energy}-\ref{fig:example:low-energy}, we next consider the theoretical  uncertainty bands 
obtained by variation of the perturbative scales. 
We perform the scale variation according to the 
$\zeta$ prescription approach in 
 \cite{Scimemi:2017etj,Scimemi:2018xaf}. This amounts to 
 varying two scales,  
 the factorization scale in the DY cross section  
 formula   
 and the small-$b$ matching scale in the OPE expansion of the 
 solution of TMD evolution equations. (The small-$b$ matching 
 scale of the CS-kernel  is  present in this 
 approach but its variation is not included  in 
 the calculation.) 
We vary these scales by factors $c$ in the range $[0.5,2]$, and take the maximum symmetrized deviation. 
The resulting bands are shown in fig.~\ref{fig:scale-vary}. 


We observe that the PDF uncertainty bands in fig.~\ref{fig:example:high-energy}  
are  comparable  or larger than  the perturbative scale variation bands in 
fig.~\ref{fig:scale-vary}. This underlines that  DY 
transverse momentum measurements in the TMD region 
are potentially useful to  place 
  constraints on  PDFs. For this purpose one needs to employ   a 
  theoretical framework capable 
  of describing the low transverse momentum region.  One could envisage 
 doing  this in  
  a resummation  framework, formulated  in terms of  collinear PDF only, or 
  in a TMD framework, in which a joined fit of both 
  PDFs and TMDPDFs will put extra constraints on the PDFs. 

\subsection{Extracted TMD distributions}

\begin{figure}
\centering
\includegraphics[width=\textwidth]{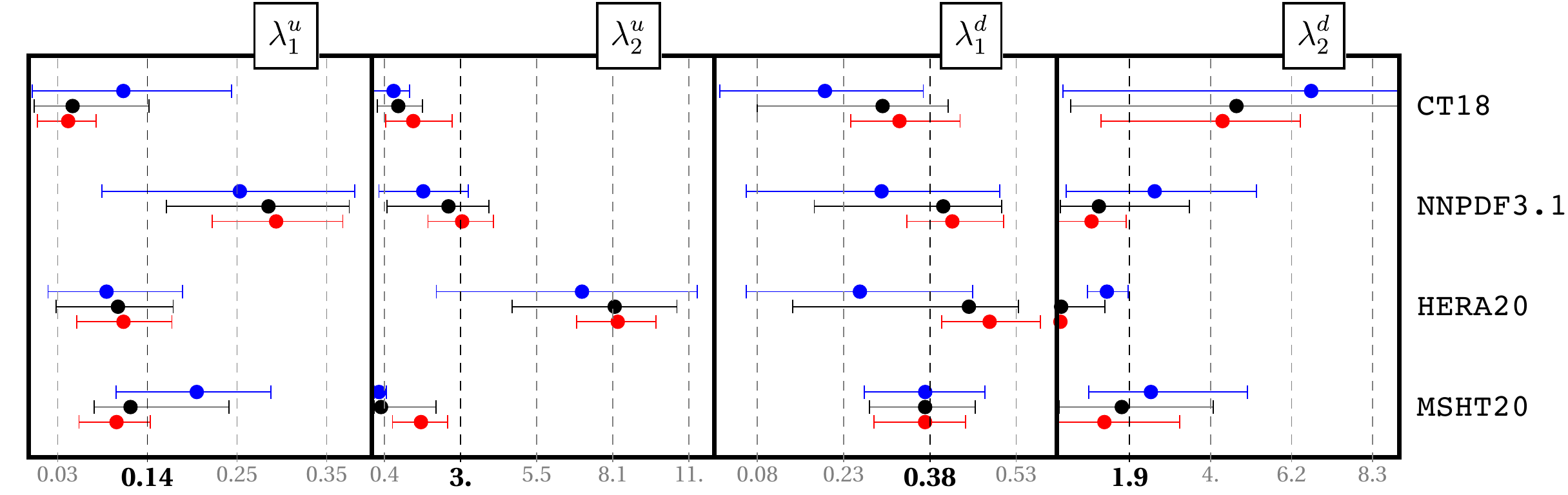}
\includegraphics[width=\textwidth]{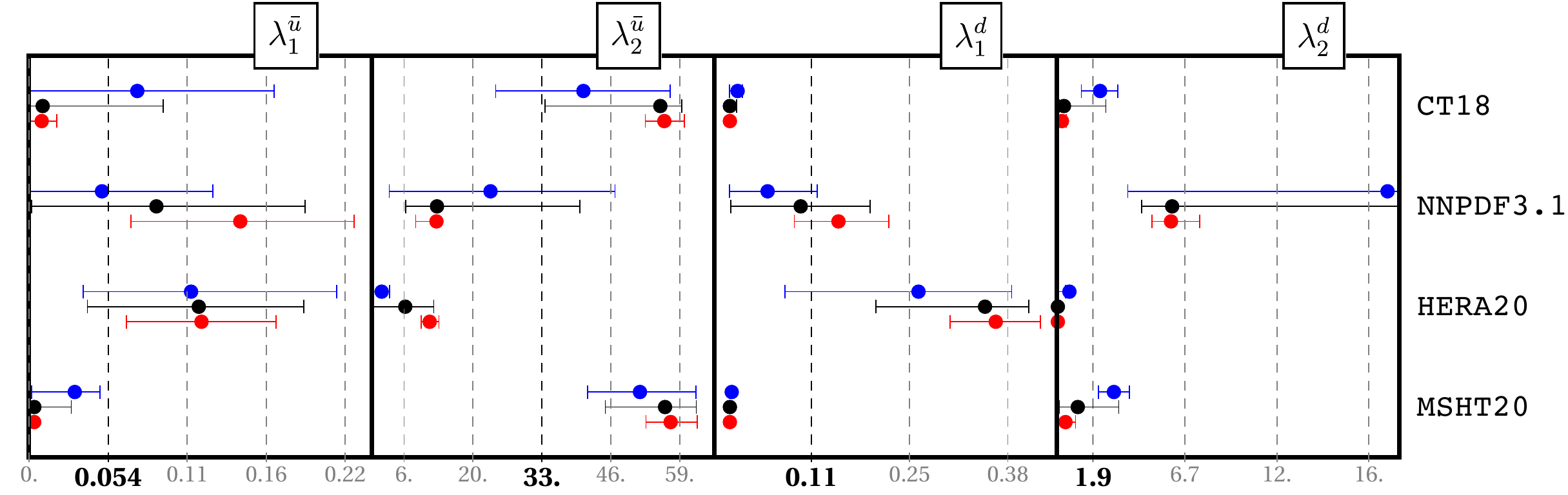}
\includegraphics[width=\textwidth]{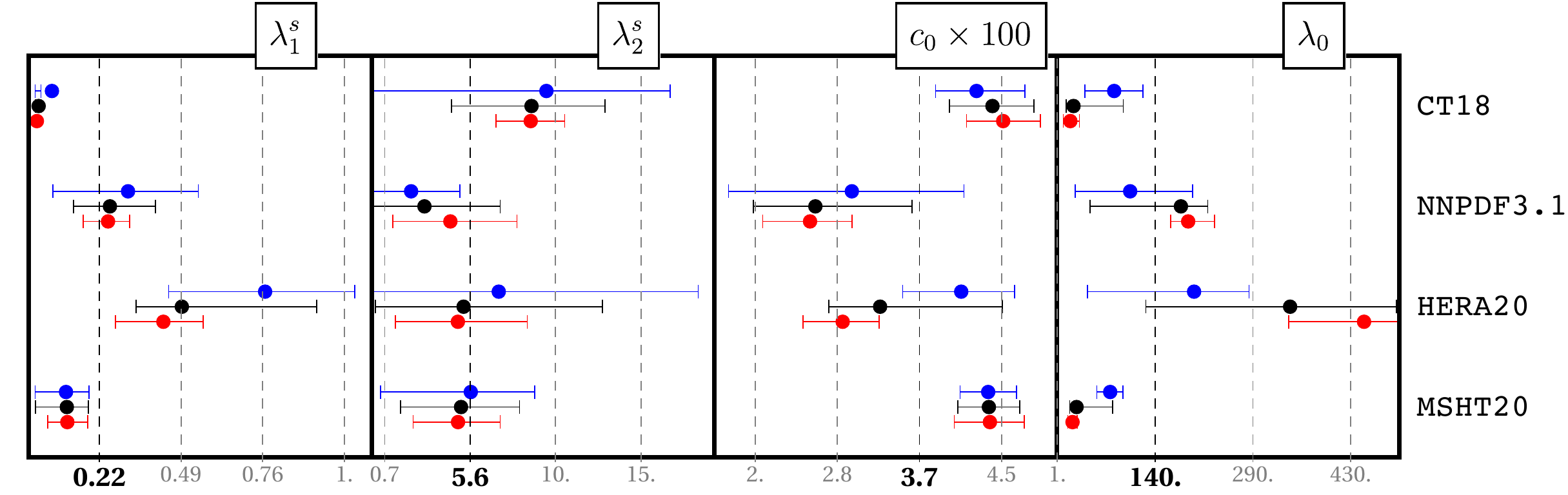}
\caption{\label{fig:Parameters}
Comparison of the parameter values. Black is the final result. Blue is the value from the fit of the {\bf PDF} case. Red is the value from the fit of the {\bf EXP} case.
}
\end{figure}

The values and error bars of the fitted parameters 
 for the TMD 
distributions and CS kernel 
are given in tab.~\ref{tab:parameters} and plotted in fig.~\ref{fig:Parameters}. There, we report the bands obtained from the estimation of the {\bf PDF} error (blue), the {\bf EXP} error (red), and the averaged result (black). The central values for the \textbf{PDF}, \textbf{EXP} and joined cases do not coincide because they are computed with the respective replicas as explained in  sec.~\ref{sec:statistical}. 

We observe  from fig.~\ref{fig:Parameters}  that the \textbf{PDF} error (blue) is generally much larger than the \textbf{EXP} error (red). That is,   the error due to a single PDF-set uncertainty is always the most significant. This aspect can be relevant for future PDF analyses and should be taken into account by future TMDPDF fits.

We also see from fig.~\ref{fig:Parameters} that there is a general agreement among parameters within error bands. However, each PDF case has a few parameters whose values deviate significantly from the rest. These are \{$\lambda_2^d, \lambda_{1,2}^s\}$ for CT18, $\{\lambda_1^u, \lambda_2^{\bar d}\}$ for NNPDF3.1, $\{\lambda_2^u,\lambda_2^{\bar u}\}$ for HERA20, and $\lambda_1^{\bar u}$ for MSHT20. This highlights the tension in the corresponding domain between PDF and TMDPDF extractions. This may also  be related to the larger values of $\chi^2$ found in flavor-independent fits of TMDPDFs.

\begin{table}[b]
\begin{center}
\renewcommand{\arraystretch}{1.1}
\begin{tabular}{|c  V{4} c V{4} c V{4} c V{4} c V{4}}
\hline
Parameter & \textbf{MSHT20}  & \textbf{HERA20}  & \textbf{NNPDF31}  & \textbf{CT18}  \\
\hline \hline
$\lambda_1^u$ & $0.12_{-0.04}^{+0.12}$ & $0.11_{-0.07}^{+0.07}$ & $0.28_{-0.10}^{+0.12}$ & $0.05_{-0.05}^{+0.09}$
\\
$\lambda_2^u$ & $0.32_{-0.22}^{+1.84}$ & $8.15_{-3.51}^{+2.09}$ & $2.58_{-2.05}^{+1.37}$ & $0.9_{-0.71}^{+0.84}$
\\\hline
$\lambda_1^d$ & $0.37_{-0.10}^{+0.09}$ & $0.44_{-0.31}^{+0.09}$ & $0.40_{-0.22}^{+0.10}$ & $0.29_{-0.22}^{+0.11}$
\\
$\lambda_2^d$ & $1.7_{-1.6}^{+2.4}$ & $0.11_{-0.11}^{+1.14}$ & $1.1_{-1.0}^{+2.4}$ & $4.7_{-4.4}^{+5.0}$
\\\hline
$\lambda_1^{\bar u}
\times 100$ & $0.37_{-0.27}^{+2.51}$ & $11.6_{-7.6}^{+7.1}$ & $8.8_{-8.6}^{+10.1}$ & $0.94_{-0.85}^{+8.14}$
\\
$\lambda_2^{\bar u}$ & $56._{-12.}^{+6.}$ & $6.5_{-6.4}^{+5.5}$ & $13._{-6.}^{+28.}$ & $56._{-22.}^{+4.}$
\\\hline
$\lambda_1^{\bar d} \times 100$ 
& $0.12_{-0.02}^{+0.18}$ & $35._{-15.}^{+6.} $ & $9.8_{-9.5}^{+9.4}$ & $0.12_{-0.02}^{+0.86}$
\\
$\lambda_2^{\bar d}$ & $1.1_{-1.0}^{+2.4}$ & $0.05_{-0.05}^{+0.09}$ & $6.1_{-1.6}^{+16.7}$ & $0.37_{-0.26}^{+2.20}$
\\\hline
$\lambda_1^{s}$ & $0.11_{-0.10}^{+0.07}$ & $0.49_{-0.15}^{+0.45}$ & $0.25_{-0.12}^{+0.15}$ & $0.012_{-0.011}^{+0.006}$
\\
$\lambda_2^{s}$ & $5.1_{-3.4}^{+3.3}$ & $5.2_{-5.0}^{+7.7}$ & $3.0_{-3.0}^{+4.3}$ & $9.1_{-4.5}^{+4.2}$
\\\hline
$\lambda_0$ & $29._{-10.}^{+53.}$ & $339._{-212.}^{+156.}$ & $181._{-133.}^{+39.}$ & $24_{-11.}^{+73.}$
\\\hline
$c_0\times 100$ & $4.36_{-0.31}^{+0.31}$ & $3.27_{-0.52}^{+1.23}$ & $2.61_{-0.61}^{+0.97}$ & $4.39_{-0.44}^{+0.41}$
\\\hline
\end{tabular}
\caption{\label{tab:parameters} The values of the NP parameters obtained in the fit.}
\end{center}
\end{table}

The fits in refs.~\cite{Bertone:2019nxa,Scimemi:2019cmh,Bacchetta:2019sam,Vladimirov:2019bfa} have found 
deficits in the predicted cross sections compared to measurements, with the deficits being small in the case of  high-energy measurements (typically, $1$-$4$\%) but 
more significant  for 
low-energy measurements (typically, $30$-$40$\%).    
 In most cases, the discrepancy in the normalization is compensated by the correlated uncertainty of the experiment (e.g., E288 measurements have 25\% uncertainty due to the beam luminosity), and thus does not significantly increase the value of $\chi^2$. Each PDF set requires its own normalization factor, e.g. for the central rapidity $Z$-boson production measured at ATLAS 8 TeV, the deficits are \{3.6, 1.4, 2.1, 6.4\}\% for \{MSHT20, HERA20, NNPDF3.1, CT18\} cases. Another example is E288 at 200 GeV with \{30, 49, 40, 39\}\%, correspondingly. 

\begin{figure}
\centering
\includegraphics[width=0.45\textwidth]{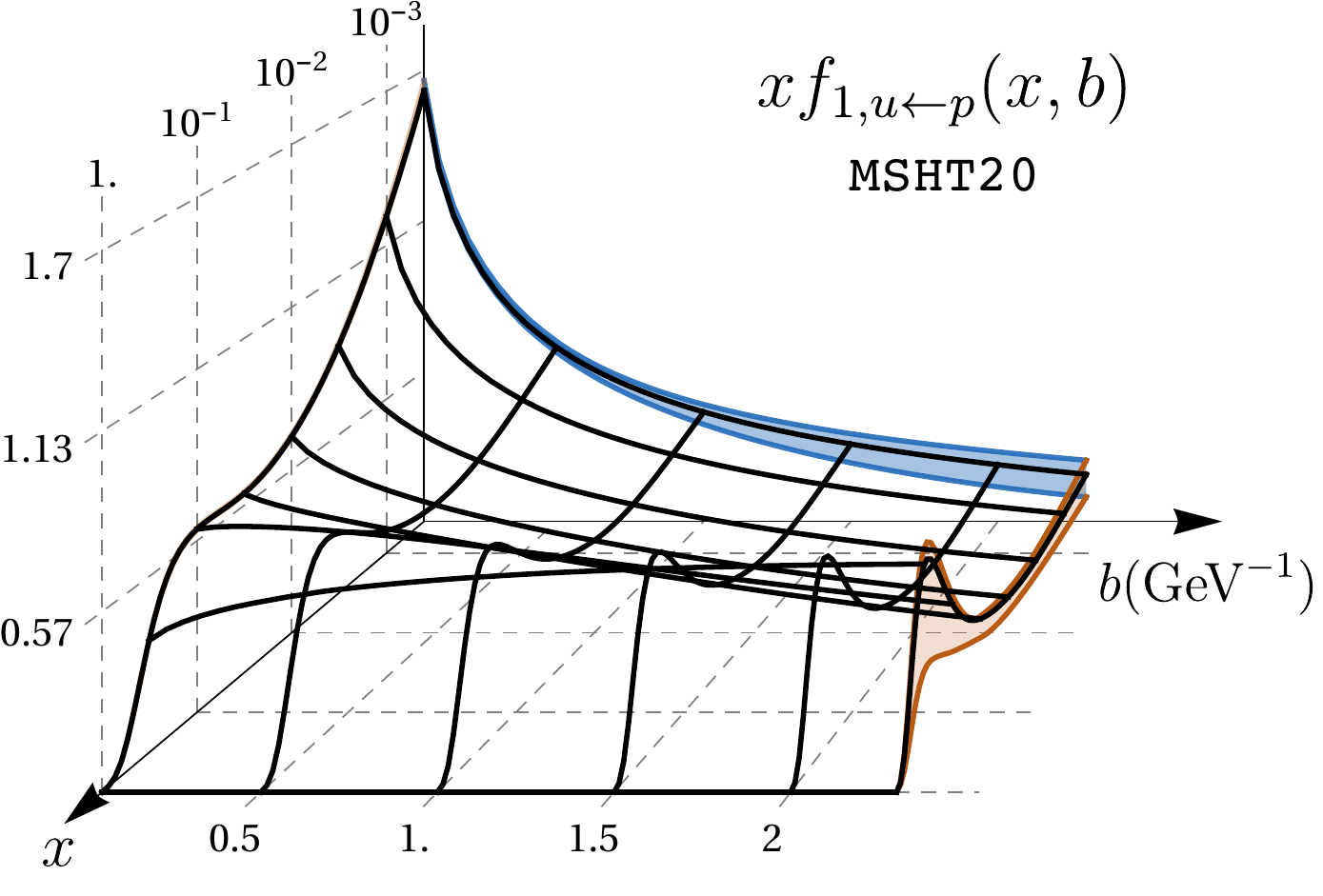}
\includegraphics[width=0.45\textwidth]{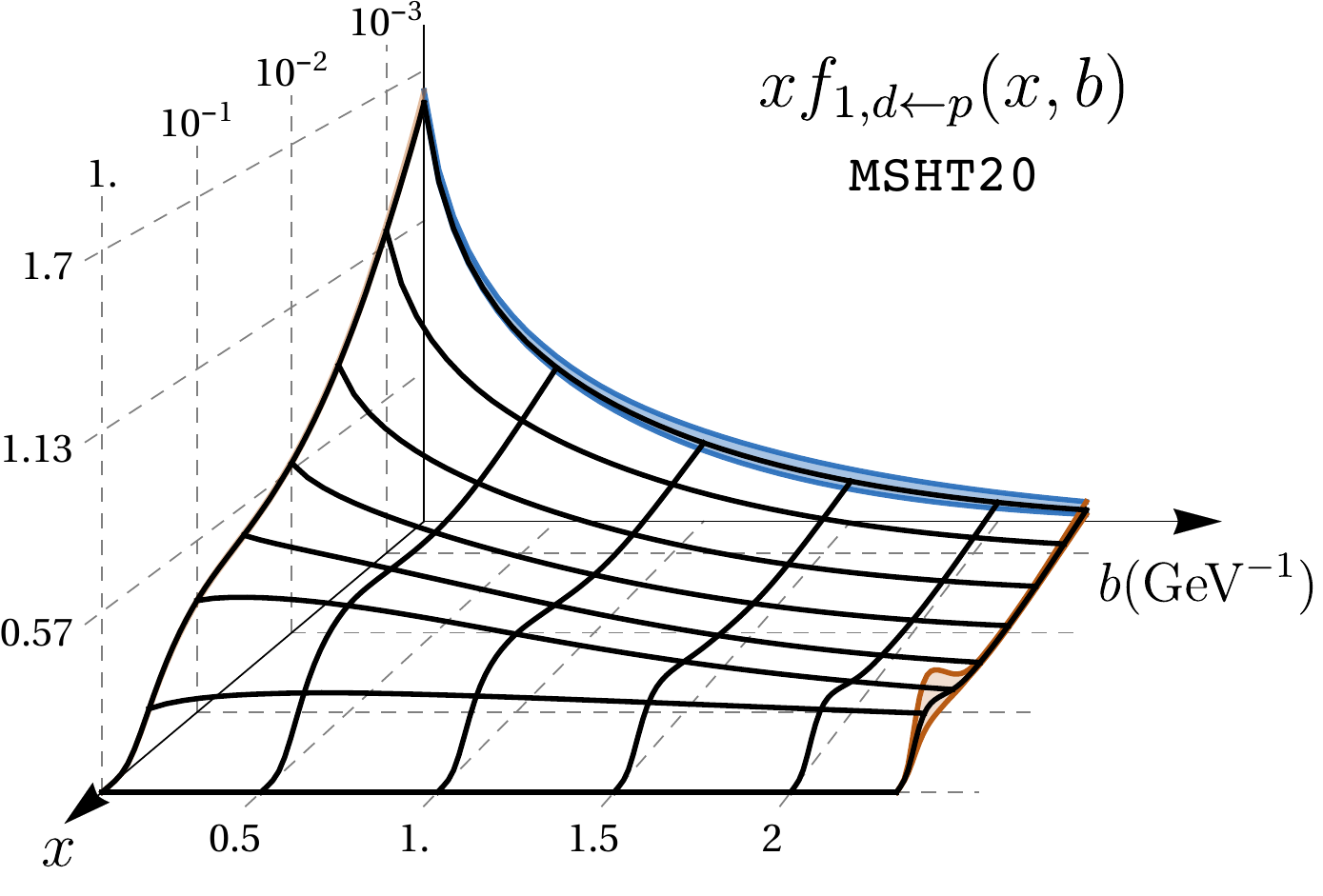}
\caption{\label{fig:sofa}
The optimal TMDPDF as a function of $(x,b)$ for $u$ and $d$ quarks with the MSHT20 PDF-input. The uncertainty is demonstrated at boundaries.
}
\end{figure}

\begin{figure}
\centering
\includegraphics[width=0.85\textwidth]{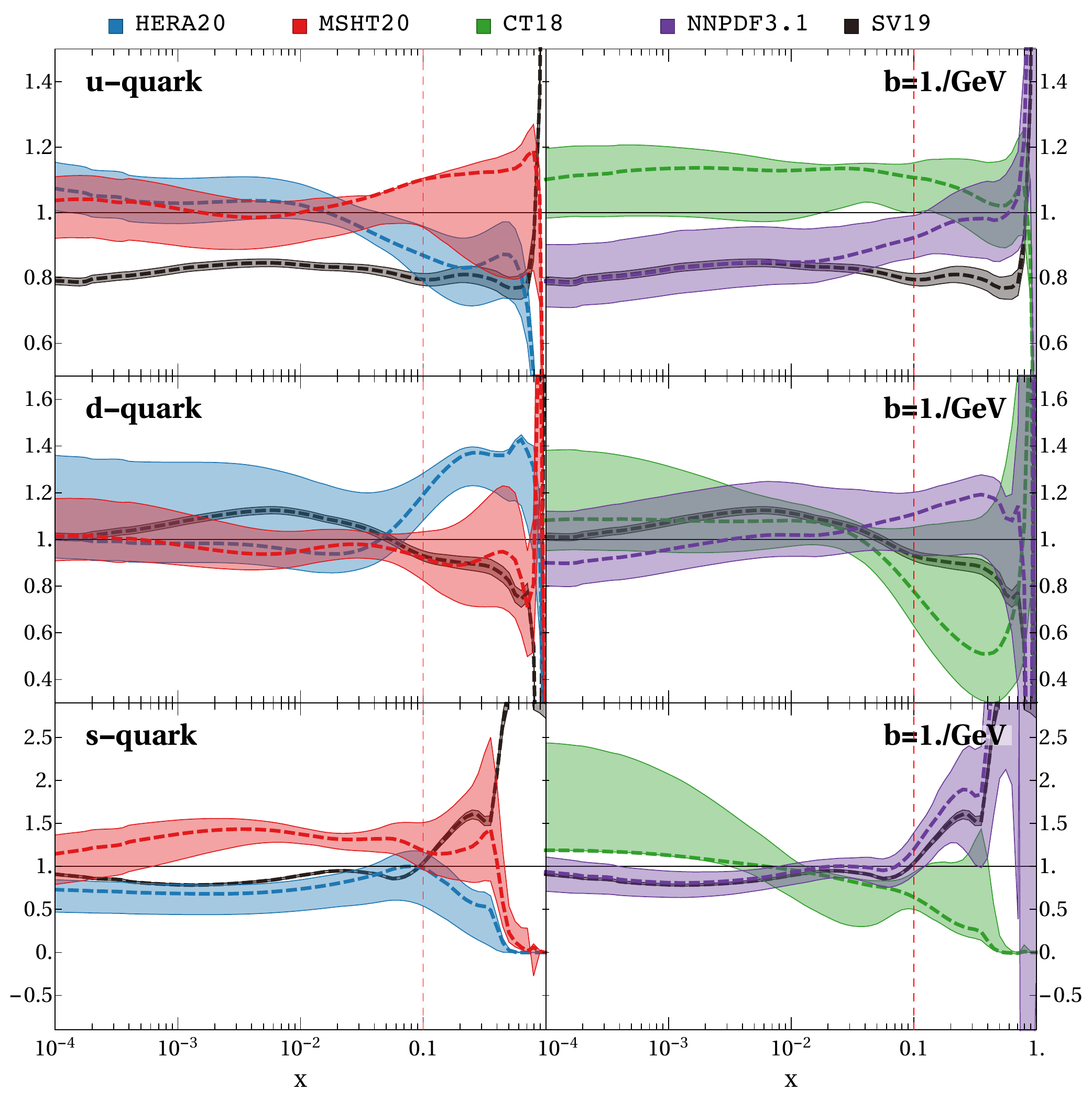}
\caption{\label{fig:compare-vs-X}
Comparison of uncertainty band for unpolarized TMDPDFs extracted with different PDFs. Here, the slice of optimal TMDPDF at $b=1$GeV$^{-1}$ is shown as the function of $x$. For convenience of presentation the plot is weighted with the central TMDPDF value averaged between different PDF cases. The red line indicates the position of slice demonstrated in fig.\ref{fig:compare-vs-B}.
}
\end{figure}

\begin{figure}
\centering
\includegraphics[width=0.85\textwidth]{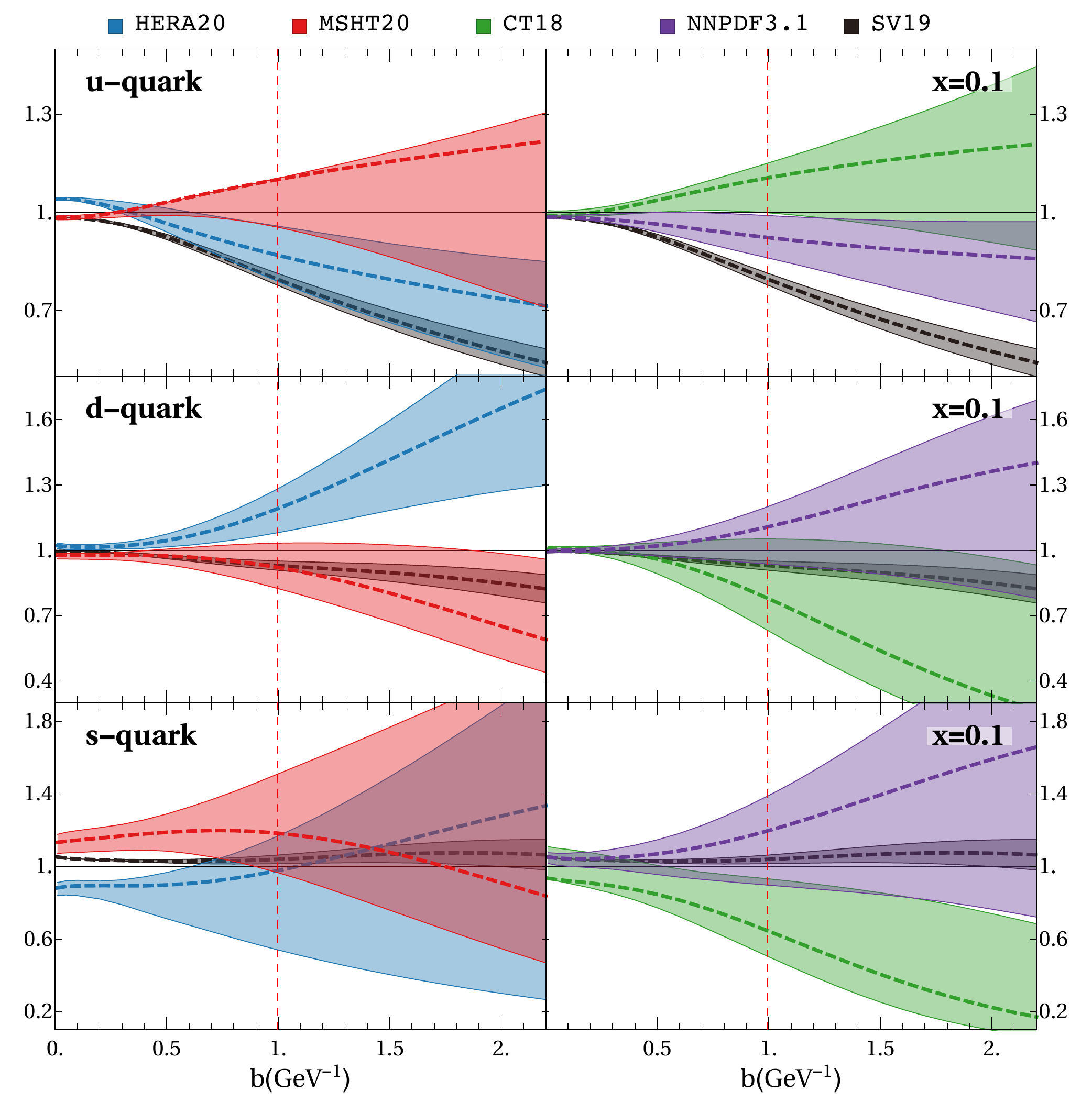}
\caption{\label{fig:compare-vs-B}
Comparison of the uncertainty band for unpolarized TMDPDFs extracted with different PDFs. Here, the slice of optimal TMDPDF at $x=0.1$ is shown as a function of $b$. For convenience of presentation the plot is weighted with the central TMDPDF value averaged between different PDF cases. The red line indicates the position of slice demonstrated in fig. \ref{fig:compare-vs-X}.
}
\end{figure}

The actual shapes of the extracted TMDPDF are summarized in figs.~\ref{fig:sofa},~\ref{fig:compare-vs-X} and \ref{fig:compare-vs-B}. We plot the 
``optimal'' \cite{Scimemi:2019cmh,Vladimirov:2019bfa} TMD distribution, that is, the  
distribution  defined according to the 
$\zeta$-prescription  \cite{Scimemi:2017etj,Scimemi:2018xaf} as the reference, scaleless TMD distribution. 
Fig.~\ref{fig:sofa} provides the overall picture as a function of 
$b$ and $x$. 
 A more detailed view is offered by the slice in $b$ in  fig.~\ref{fig:compare-vs-X}  and the slice in $x$  in fig.~\ref{fig:compare-vs-B}, showing the TMDPDF for each 
PDF set divided by the averaged central value of all PDF cases. The size of the uncertainty varies strongly from the low-$b$ region, where we have a good knowledge of the perturbative expansion, to the  non-perturbative high-$b$ region (fig.~\ref{fig:compare-vs-X}). For $b\geq 2$ GeV$^{-1}$ the relative uncertainty on the TMD distributions is not less than 60-80$\%$.  
For comparison,  figs.~\ref{fig:compare-vs-X} and \ref{fig:compare-vs-B} also show the uncertainty band 
obtained in the fit \cite{Scimemi:2019cmh}, labelled SV19. 
One of the main outcomes of the present work is that, 
compared to previous DY and SIDIS fits such as \cite{Scimemi:2019cmh}, 
the TMD uncertainty obtained in this paper 
is about 4-5 times larger, 
as a result of the improved analysis framework  
taking into 
account the propagation of collinear PDF uncertainties to the 
TMD extraction and the flavor dependence of the TMD profile. 
Furthermore, note  that  
the present extraction has a non-zero uncertainty band also 
at $b=0$, which was instead forbidden by construction in 
all previous studies. 


\begin{figure}
\centering
\includegraphics[width=0.65\textwidth]{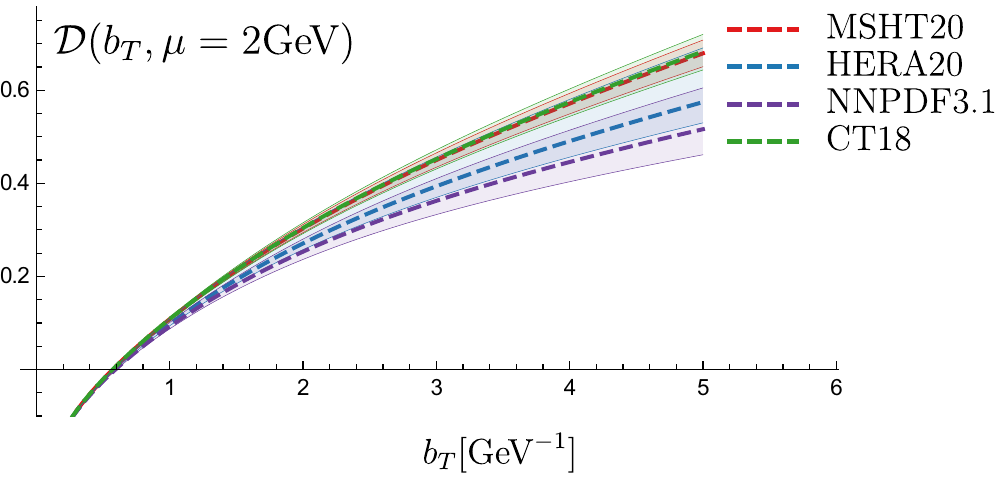}
\caption{\label{fig:RAD}
The CS-kernel as a function of $b$ extracted with different PDF inputs.
}
\end{figure}

The results presented in this section indicate that, compared to previous 
DY and SIDIS fits,   
the  approach of the present paper 
leads to a more reliable 
estimate of the TMD uncertainties, to a reduced spread 
in the $\chi^2$ distribution for each PDF set,  
and to a better agreement between different PDF sets. 
Nevertheless, we see from figs.~\ref{fig:compare-vs-X} and \ref{fig:compare-vs-B} that 
the TMDPDFs extracted with different PDFs 
still display significant differences. 
A similar remark applies to the CS kernel: 
results for the CS kernel extracted with different PDFs 
are shown in fig.~\ref{fig:RAD}. 

\begin{figure}
\centering
\includegraphics[width=0.85\textwidth]{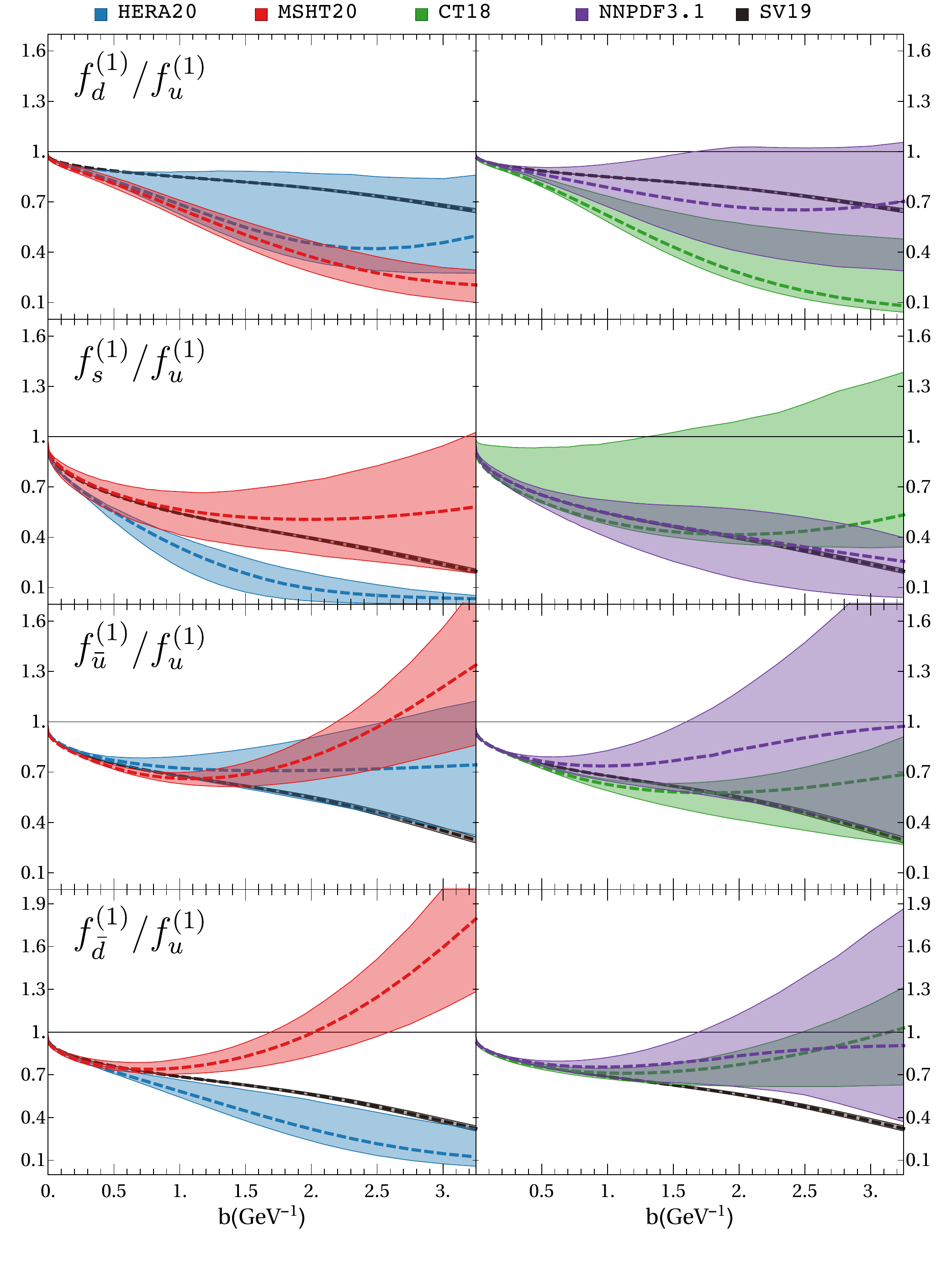}
\caption{\label{fig:compare-M1}
The ratio of the first Mellin moments of unpolarized TMDPDF (\ref{def:moments}) as a function of $b$ for different PDF cases. 
}
\end{figure}

To further investigate  the variation of TMDPDFs with the 
PDF set, we turn to 
Mellin moments of the TMDPDFs. 
We consider the following quantities  
\begin{eqnarray}\label{def:moments}
f^{(n)}_{f}(b)=\int_{x_{\text{min}}}^1 dx\, x^{n-1} f_{1,f\ot h}(x,b),
\end{eqnarray}
where we set $x_{\text{min}}=10^{-5}$. One may expect that the dependence on the collinear PDF is reduced, for $n=1$ and $n=2$, due to momentum sum rules fulfilled 
by each PDF set. Note however that    $f_{\text{NP}}$ in  eq.~(\ref{def:fNP}) contains a (weak) dependence on 
$x$. At any rate, the investigation of moments is significant 
because  ratios for different flavours can be measured independently using lattice methods, see e.g. refs.~\cite{Musch:2011er,Schlemmer:2021aij}. 
The ratios of the first moments for the light quarks with respect to the $u$-quark are shown in fig.~\ref{fig:compare-M1}. We see that, with respect to point-to-point comparisons, they display a better global agreement. A similar behaviour appears for 
the ratios of second and third moments. The corresponding plots are given in the supplementary material. 

We conclude this subsection by noting that the results for  
unpolarized TMD distributions presented above  will  also be 
 important  for the phenomenology of spin-dependent TMD distributions, where the unpolarized cross-section serves as the normalization for the measurements of polarization asymmetries and angular distributions. For example, the uncertainty band from the 
 fit \cite{Scimemi:2019cmh} resulted into 10-15\% of the total uncertainty band for the Sivers function \cite{Bury:2020vhj}. 
 With the results of the present paper one can, on  one hand,  
 expect an extra $\sim 50\%$ uncertainty due to the  inclusion of the PDF uncertainty. On the other hand, one can also expect a better agreement with experiments, due to a proper flavor dependence 
 in the unpolarized sector. 

\subsection{Predictions for $W$-boson production}
\label{sec:W}

\begin{table}[h]
\begin{center}
\begin{tabular}{|l|c|c|c|c|c|c|c|}\hline
& CDF \cite{Abe:1991rk}  &  D0  \cite{Abbott:1998jy} & \multicolumn{3}{c|}{ATLAS \cite{Aad:2011fp}} & \multicolumn{2}{c|}{CMS \cite{Khachatryan:2016nbe}} \\
 & & & $e$-channel & $\mu$-channel & combined & $e$-channel & $\mu$-channel
\\\hline
Points& 10 &10&2& 2&2&4&4\\\hline\hline
MSHT20 & 0.66& 1.8&  2.9& 1.6& 2.5&  8.1& 32. 
\\\hline
HERA20 & 0.68& 2.0&  1.7& 0.6& 1.2& 5.9& 23.
\\\hline
NNPDF31 & 0.70& 1.8& 2.5& 1.3& 2.0& 7.6& 30.
\\\hline
CT18  & 0.71& 1.9& 1.7& 0.7& 1.2& 7.1& 26.
\\\hline
\end{tabular}
\end{center}
\caption{\label{tab:WPDFs2} $\chi^2/N$  for differential $W$ production using the current extraction of TMD  and theoretical errors (
including PDF error coming from 1000 replicas).}
\end{table}

The flavor dependence of the TMD distributions could significantly impact the description of processes mediated by the $W$-boson  \cite{Signori:2013mda,Bacchetta:2018lna,Signori:2016lvd}. In this work we have demonstrated that flavor dependence is essential to  improve the agreement between theory and data. Including  $W$-boson production data in future studies of unpolarized TMDPDFs will thus be very 
important. 

Since it significantly slows down the fitting procedure, we have not included $W$-boson production data in 
this study. Instead, we compare the prediction made with the present extraction to the one made in  \cite{Gutierrez-Reyes:2020ouu}.
The computation is performed with the \texttt{artemide} code, appropriately adapted for the computation of the transverse-mass-differential 
cross-section as described in ref.~\cite{Gutierrez-Reyes:2020ouu}. The values of the $\chi^2$ for comparison with the data by Tevatron and LHC experiments \cite{Abbott:1998jy,Abe:1991rk,Khachatryan:2016nbe,Aad:2011fp} are presented in tab. \ref{tab:WPDFs2}. We find a small general improvement compared to 
ref.~\cite{Gutierrez-Reyes:2020ouu}, which is expected because most of $W$-boson production data belong to the resummation region. 

We see that the $\chi^2$ values for the CMS measurement are  large in comparison with the other data considered. This may be understood 
by noting that this measurement is integrated over the full range of the dilepton mass, and the integration range covers low-mass regions in which power corrections in $q_T/Q$ are non-negligible. 
 As a result, the theory prediction deviates from the measurement starting from smaller values of $q_T$, producing larger  $\chi^2$-values. This effect is partially canceled in the
  cross-section ratios $W^-/W^+$ and $Z/W$, which show a substantial agreement between theory and the CMS measurements. Additional plots on this are presented in the 
  supplementary material.

\begin{figure}[t]
\centering
\includegraphics[width=0.99\textwidth]{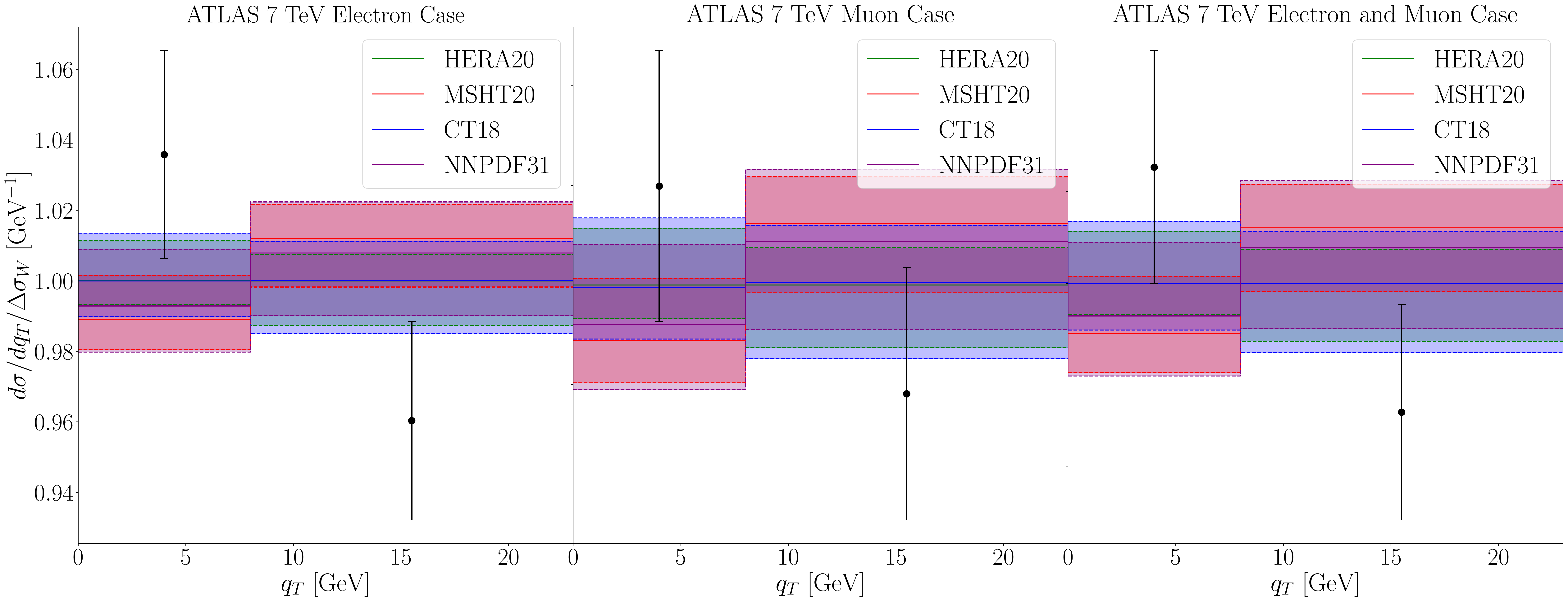}
\caption{\label{fig:Z-ATLAS} Comparison of the differential cross section for W-boson production measured by ATLAS at 7 TeV with the theory prediction using different PDFs and normalized to HERA20 case.}
\end{figure}

\section{Conclusions}
\label{sec:conclusions}

In this work 
we have performed the first quantitative 
study of the influence of  PDF on 
fits to DY transverse momentum measurements 
based on TMD factorization.  
We have referred to this as 
the PDF bias issue, arising from the fact that 
an OPE is applied to the TMD operator and 
an ansatz is made for the TMD 
distribution in terms of collinear PDFs. We have 
highlighted the quantitative 
significance of this issue in unpolarized TMD fits. 
We have stressed that this is relevant also in polarized TMD analyses, 
for which  unpolarized cross sections are used to 
normalize 
angular distributions and asymmetries. 

We have carried out a Bayesian procedure to propagate PDF uncertainties to the extraction of TMD parton distributions. 
We have examined four PDF sets (MSHT20, NNPDF31, CT18, HERA20),  representative of different NNLO PDF methodologies,  
and we have performed a TMD determination from DY data including  for the first time both experimental 
and PDF uncertainties. We have found that the PDF uncertainties  are larger than the DY experimental uncertainties 
for all values of $b$ (or $p_T$). As a result of the improved analysis framework, we have obtained 
reliable estimates of TMD uncertainties,  with the size of TMD error bands being significantly increased with respect to previous TMD fits which do not perform the full PDF bias analysis. 

We have included for the first time flavor dependence in the non-perturbative TMD distribution  $f_{\text{NP}}^f$.  Previous fits include flavor dependence in the collinear PDF only. We have found that flavor-dependent TMD profiles  
reduce the spread in $\chi^2$ distributions for each PDF set, improving the agreement between data and theory, and 
help obtain more consistent results among different PDF sets.

The results of this paper indicate that including   collinear PDF uncertainties in TMD extractions and taking into account the flavor dependence of  NP TMD distributions are both  essential to obtain reliable TMD determinations from DY (and SIDIS) transverse momentum data. Future phenomenological studies,  which incorporate these features with more powerful  computational and statistical tools than those used here, are warranted. 

\acknowledgments This study was supported by Deutsche Forschungsgemeinschaft (DFG) through the research Unit FOR 2926, “Next Generation pQCD for Hadron Structure: Preparing for the EIC”, project number 430915485. I.S. is supported by the Spanish Ministry grant PID2019-106080GB-C21. 
This project has received funding from the European Union Horizon 2020 research and innovation program under grant agreement Num. 824093 (STRONG-2020). S.L.G. is supported by the Austrian Science Fund FWF under the Doctoral Program W1252-N27 Particles
and Interactions.

\appendix
\section{Appendix}
\label{sec:pdfbias} 

In this appendix we provide a few more  
details and checks on the fits 
performed in sec.~\ref{sec:results}. 

\begin{figure}
\centering
\includegraphics[width=0.5\textwidth]{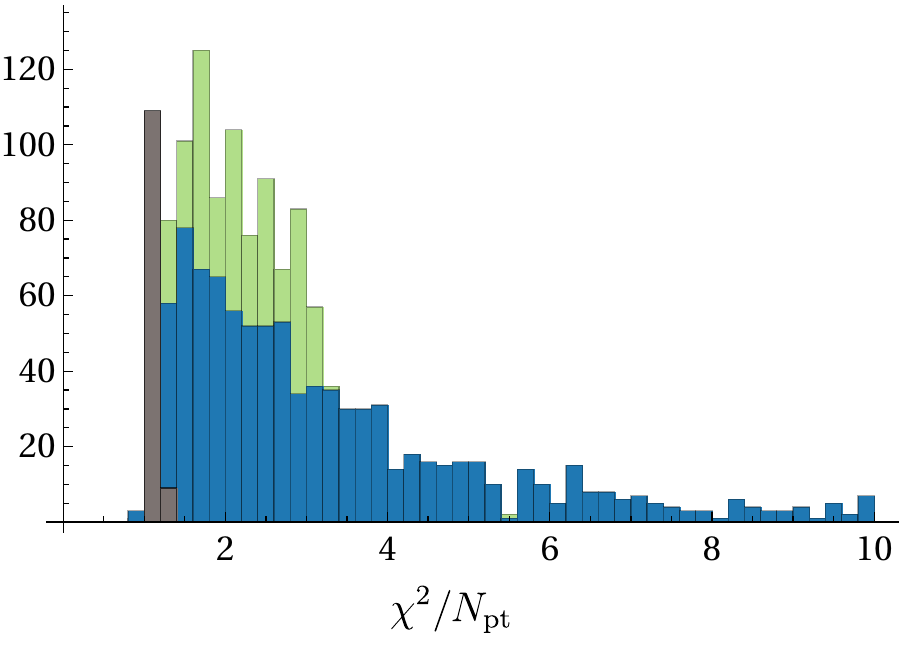}
\caption{\label{fig:SV19-3} 
The same as fig.~\ref{fig:chi2} in the case of a flavor independent $f_{\text{NP}}$ with the PDF set NNPDF3.1. 
The green histogram is obtained fitting $f_{\text{NP}}$ for each replica and the blue one is obtained keeping $f_{\text{NP}}$ fixed as from the fit of central replica. The grey histogram is the {\bf EXP} error (see definition in sec.\ref{sec:EXP+PDF}).}
\end{figure}

In fig.~\ref{fig:chi2} we have presented our results 
for  
the  distribution of $\chi^2$ values  among the {\bf PDF} 
and {\bf EXP} replicas, and we have observed that 
the $\chi^2$ spread is much  reduced with respect to 
previous fits in the literature, thanks to the  
 flavor-dependent profile used in the present work 
for the NP TMD distributions $f_{\rm NP}^f$.  
Here we illustrate  the role of the flavor dependence 
explicitly 
by reporting the result for the $\chi^2$ distribution 
which we obtain by repeating the calculation of 
fig.~\ref{fig:chi2} but replacing the flavor-dependent 
$f_{\rm NP}^f$ model of eq.~(\ref{def:fNP}) 
with the flavor-independent model 
of ref.~\cite{Scimemi:2019cmh}. 
The result is shown in fig.~\ref{fig:SV19-3} 
for the case of the 
\texttt{NNPDF3.1} PDF set \cite{Ball:2017nwa}.

The result in fig.~\ref{fig:SV19-3}   is to be compared with the third panel 
in fig.~\ref{fig:chi2}. We see that, in contrast to the 
$\chi^2$ distribution found in fig.~\ref{fig:chi2},  the 
 distribution of $\chi^2$-values over PDF replicas 
in fig.~\ref{fig:SV19-3} shows an unsatisfactorily broad 
shape, with about $64\%$ of the replicas having $\chi^2/N_{\text{pt}}>2$. We have checked 
that the unsatisfactory 
$\chi^2$-values for the subset of replicas are not due to a single problematic measurement but rather they are common to all 
data.

 We have also checked that the issues described above are not specific to the \texttt{NNPDF3.1} PDF set \cite{Ball:2017nwa}. We performed similar tests using \texttt{HERA20} \cite{Abramowicz:2015mha} ($\chi^2_0/N_{\text{pt}}=0.97$), \texttt{MMHT14} \cite{Harland-Lang:2014zoa} ($\chi^2_0/N_{\text{pt}}=1.34$), \texttt{CT14} \cite{Dulat:2015mca} ($\chi^2_0/N_{\text{pt}}=1.59$), \texttt{PDF4LHC15} \cite{Butterworth:2015oua} ($\chi^2_0/N_{\text{pt}}=1.53$), \texttt{MSHT20} \cite{Bailey:2020ooq} ($\chi^2_0/N_{\text{pt}}=1.25$), \texttt{CT18} \cite{Hou:2019efy} ($\chi^2_0/N_{\text{pt}}=1.26$), and \texttt{CJ15nlo} \cite{Accardi:2016qay} ($\chi^2_0/N_{\text{pt}}=1.82$),  where $\chi^2_0$ is the $\chi^2$-value for the central PDF replica. All these PDF sets are characterized by the same issues as \texttt{NNPDF3.1}.  
  This confirms 
 that the essential element at the origin of the difference between the $\chi^2$ distributions in 
  fig.~\ref{fig:chi2} and  fig.~\ref{fig:SV19-3}  
is  the  flavor dependence of the NP TMD distributions  $f_{\rm NP}^f$.

\begin{figure}
\centering
\includegraphics[width=0.45\textwidth]{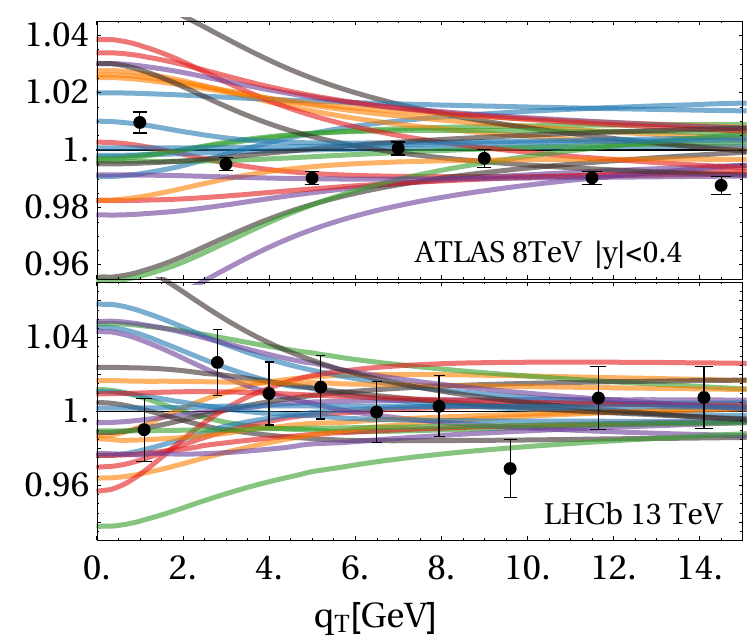}
\caption{\label{fig:SV19-2} 
Examples of prediction using different PDF replicas at the same $f_{\text{NP}}$. 
}
\end{figure} 

We next discuss the effect of different PDF replicas on the shape of the predictions for the transverse momentum distribution. 
One might wonder whether the change in PDF replicas results into an effect primarily on the normalization but not on the $q_T$ shape of the predictions. 
In fig.~\ref{fig:SV19-2} we illustrate that this is not the case.  That is, fig.~\ref{fig:SV19-2} indicates that 
the large spread in the $\chi^2$ distribution observed above is due to  different PDF replicas  inducing  different $q_T$-shapes of predictions. 
 The variety of shapes is a consequence of  the structure  of the convolution within OPE, which correlates the $b$ and  $x$ dependences.


\bibliographystyle{JHEP}
\normalbaselines 
\bibliography{TMD_ref}

\providecommand{\href}[2]{#2}\begingroup\raggedright\begin{thebibliography}{100}

\bibitem{Kovarik:2019xvh}
K.~Kova\v{r}\'\i{}k, P.~M. Nadolsky and D.~E. Soper, \emph{{Hadronic structure
  in high-energy collisions}},
  \href{http://dx.doi.org/10.1103/RevModPhys.92.045003}{\emph{Rev. Mod. Phys.}
  {\bf 92} (2020) 045003}, [\href{http://arxiv.org/abs/1905.06957}{{\tt
  1905.06957}}].

\bibitem{Angeles-Martinez:2015sea}
R.~Angeles-Martinez et~al., \emph{{Transverse Momentum Dependent (TMD) parton
  distribution functions: status and prospects}},
  \href{http://dx.doi.org/10.5506/APhysPolB.46.2501}{\emph{Acta Phys. Polon.}
  {\bf B46} (2015) 2501--2534}, [\href{http://arxiv.org/abs/1507.05267}{{\tt
  1507.05267}}].

\bibitem{Aybat:2011zv}
S.~M. Aybat and T.~C. Rogers, \emph{{TMD Parton Distribution and Fragmentation
  Functions with QCD Evolution}},
  \href{http://dx.doi.org/10.1103/PhysRevD.83.114042}{\emph{Phys. Rev.} {\bf
  D83} (2011) 114042}, [\href{http://arxiv.org/abs/1101.5057}{{\tt
  1101.5057}}].

\bibitem{Scimemi:2019cmh}
I.~Scimemi and A.~Vladimirov, \emph{{Non-perturbative structure of
  semi-inclusive deep-inelastic and Drell-Yan scattering at small transverse
  momentum}}, \href{http://dx.doi.org/10.1007/JHEP06(2020)137}{\emph{JHEP} {\bf
  06} (2020) 137}, [\href{http://arxiv.org/abs/1912.06532}{{\tt 1912.06532}}].

\bibitem{Bertone:2019nxa}
V.~Bertone, I.~Scimemi and A.~Vladimirov, \emph{{Extraction of unpolarized
  quark transverse momentum dependent parton distributions from
  Drell-Yan/Z-boson production}},
  \href{http://dx.doi.org/10.1007/JHEP06(2019)028}{\emph{JHEP} {\bf 06} (2019)
  028}, [\href{http://arxiv.org/abs/1902.08474}{{\tt 1902.08474}}].

\bibitem{Scimemi:2017etj}
I.~Scimemi and A.~Vladimirov, \emph{{Analysis of vector boson production within
  TMD factorization}},
  \href{http://dx.doi.org/10.1140/epjc/s10052-018-5557-y}{\emph{Eur. Phys. J.}
  {\bf C78} (2018) 89}, [\href{http://arxiv.org/abs/1706.01473}{{\tt
  1706.01473}}].

\bibitem{Bacchetta:2019sam}
A.~Bacchetta, V.~Bertone, C.~Bissolotti, G.~Bozzi, F.~Delcarro, F.~Piacenza
  et~al., \emph{{Transverse-momentum-dependent parton distributions up to
  N$^{3}$LL from Drell-Yan data}},
  \href{http://dx.doi.org/10.1007/JHEP07(2020)117}{\emph{JHEP} {\bf 07} (2020)
  117}, [\href{http://arxiv.org/abs/1912.07550}{{\tt 1912.07550}}].

\bibitem{Bacchetta:2017gcc}
A.~Bacchetta, F.~Delcarro, C.~Pisano, M.~Radici and A.~Signori,
  \emph{{Extraction of partonic transverse momentum distributions from
  semi-inclusive deep-inelastic scattering, Drell-Yan and Z-boson production}},
  \href{http://dx.doi.org/10.1007/JHEP06(2017)081}{\emph{JHEP} {\bf 06} (2017)
  081}, [\href{http://arxiv.org/abs/1703.10157}{{\tt 1703.10157}}].

\bibitem{Hautmann:2013tba}
F.~Hautmann and H.~Jung, \emph{{Transverse momentum dependent gluon density
  from DIS precision data}},
  \href{http://dx.doi.org/10.1016/j.nuclphysb.2014.03.014}{\emph{Nucl. Phys. B}
  {\bf 883} (2014) 1}, [\href{http://arxiv.org/abs/1312.7875}{{\tt
  1312.7875}}].

\bibitem{Kotko:2017oxg}
P.~Kotko, K.~Kutak, S.~Sapeta, A.~M. Stasto and M.~Strikman, \emph{{Estimating
  nonlinear effects in forward dijet production in ultra-peripheral heavy ion
  collisions at the LHC}},
  \href{http://dx.doi.org/10.1140/epjc/s10052-017-4906-6}{\emph{Eur. Phys. J.
  C} {\bf 77} (2017) 353}, [\href{http://arxiv.org/abs/1702.03063}{{\tt
  1702.03063}}].

\bibitem{BermudezMartinez:2018fsv}
A.~Bermudez~Martinez, P.~Connor, H.~Jung, A.~Lelek, R.~\v{Z}leb\v{c}\'\i{}k,
  F.~Hautmann et~al., \emph{{Collinear and TMD parton densities from fits to
  precision DIS measurements in the parton branching method}},
  \href{http://dx.doi.org/10.1103/PhysRevD.99.074008}{\emph{Phys. Rev. D} {\bf
  99} (2019) 074008}, [\href{http://arxiv.org/abs/1804.11152}{{\tt
  1804.11152}}].

\bibitem{Abdulov:2021ivr}
N.~A. Abdulov et~al., \emph{{TMDlib2 and TMDplotter: a platform for 3D hadron
  structure studies}},
  \href{http://dx.doi.org/10.1140/epjc/s10052-021-09508-8}{\emph{Eur. Phys. J.
  C} {\bf 81} (2021) 752}, [\href{http://arxiv.org/abs/2103.09741}{{\tt
  2103.09741}}].

\bibitem{Hautmann:2014kza}
F.~Hautmann, H.~Jung, M.~Kramer, P.~J. Mulders, E.~R. Nocera, T.~C. Rogers
  et~al., \emph{{TMDlib and TMDplotter: library and plotting tools for
  transverse-momentum-dependent parton distributions}},
  \href{http://dx.doi.org/10.1140/epjc/s10052-014-3220-9}{\emph{Eur. Phys. J.}
  {\bf C74} (2014) 3220}, [\href{http://arxiv.org/abs/1408.3015}{{\tt
  1408.3015}}].

\bibitem{Lhcew:2022pt}
A.~Apyan, D.~Froidevaux et~al., \emph{{LHC Electroweak Working Group: W/Z
  transverse momentum benchmarking}}, {\emph{CERN} (2021) }.

\bibitem{Camarda:2019zyx}
S.~Camarda et~al., \emph{{DYTurbo: Fast predictions for Drell-Yan processes}},
  \href{http://dx.doi.org/10.1140/epjc/s10052-020-7757-5}{\emph{Eur. Phys. J.
  C} {\bf 80} (2020) 251}, [\href{http://arxiv.org/abs/1910.07049}{{\tt
  1910.07049}}].

\bibitem{Camarda:2021ict}
S.~Camarda, L.~Cieri and G.~Ferrera, \emph{{Drell\textendash{}Yan lepton-pair
  production: qT resummation at N3LL accuracy and fiducial cross sections at
  N3LO}}, \href{http://dx.doi.org/10.1103/PhysRevD.104.L111503}{\emph{Phys.
  Rev. D} {\bf 104} (2021) L111503},
  [\href{http://arxiv.org/abs/2103.04974}{{\tt 2103.04974}}].

\bibitem{Coradeschi:2017zzw}
F.~Coradeschi and T.~Cridge, \emph{{reSolve \textemdash{} A transverse momentum
  resummation tool}},
  \href{http://dx.doi.org/10.1016/j.cpc.2018.11.024}{\emph{Comput. Phys.
  Commun.} {\bf 238} (2019) 262}, [\href{http://arxiv.org/abs/1711.02083}{{\tt
  1711.02083}}].

\bibitem{Accomando:2019ahs}
E.~Accomando et~al., \emph{{Production of Z'-boson resonances with large width
  at the LHC}},
  \href{http://dx.doi.org/10.1016/j.physletb.2020.135293}{\emph{Phys. Lett. B}
  {\bf 803} (2020) 135293}, [\href{http://arxiv.org/abs/1910.13759}{{\tt
  1910.13759}}].

\bibitem{Bizon:2018foh}
W.~Bizon, X.~Chen, A.~Gehrmann-De~Ridder, T.~Gehrmann, N.~Glover, A.~Huss
  et~al., \emph{{Fiducial distributions in Higgs and Drell-Yan production at
  N$^{3}$LL+NNLO}},
  \href{http://dx.doi.org/10.1007/JHEP12(2018)132}{\emph{JHEP} {\bf 12} (2018)
  132}, [\href{http://arxiv.org/abs/1805.05916}{{\tt 1805.05916}}].

\bibitem{Chen:2022cgv}
X.~Chen, T.~Gehrmann, E.~W.~N. Glover, A.~Huss, P.~F. Monni, E.~Re et~al.,
  \emph{{Third-Order Fiducial Predictions for Drell-Yan Production at the
  LHC}}, \href{http://dx.doi.org/10.1103/PhysRevLett.128.252001}{\emph{Phys.
  Rev. Lett.} {\bf 128} (2022) 252001},
  [\href{http://arxiv.org/abs/2203.01565}{{\tt 2203.01565}}].

\bibitem{Ebert:2020dfc}
M.~A. Ebert, J.~K.~L. Michel, I.~W. Stewart and F.~J. Tackmann,
  \emph{{Drell-Yan $q_{T}$ resummation of fiducial power corrections at
  N$^{3}$LL}}, \href{http://dx.doi.org/10.1007/JHEP04(2021)102}{\emph{JHEP}
  {\bf 04} (2021) 102}, [\href{http://arxiv.org/abs/2006.11382}{{\tt
  2006.11382}}].

\bibitem{Ebert:2016gcn}
M.~A. Ebert and F.~J. Tackmann, \emph{{Resummation of Transverse Momentum
  Distributions in Distribution Space}},
  \href{http://dx.doi.org/10.1007/JHEP02(2017)110}{\emph{JHEP} {\bf 02} (2017)
  110}, [\href{http://arxiv.org/abs/1611.08610}{{\tt 1611.08610}}].

\bibitem{Becher:2020ugp}
T.~Becher and T.~Neumann, \emph{{Fiducial $q_T$ resummation of color-singlet
  processes at N$^3$LL+NNLO}},
  \href{http://dx.doi.org/10.1007/JHEP03(2021)199}{\emph{JHEP} {\bf 03} (2021)
  199}, [\href{http://arxiv.org/abs/2009.11437}{{\tt 2009.11437}}].

\bibitem{Vladimirov:2019bfa}
A.~Vladimirov, \emph{{Pion-induced Drell-Yan processes within TMD
  factorization}},  \href{http://arxiv.org/abs/1907.10356}{{\tt 1907.10356}}.

\bibitem{Bury:2020vhj}
M.~Bury, A.~Prokudin and A.~Vladimirov, \emph{{Extraction of the Sivers
  Function from SIDIS, Drell-Yan, and $W^{\pm}/Z$ Data at
  Next-to-Next-to-Next-to Leading Order}},
  \href{http://dx.doi.org/10.1103/PhysRevLett.126.112002}{\emph{Phys. Rev.
  Lett.} {\bf 126} (2021) 112002}, [\href{http://arxiv.org/abs/2012.05135}{{\tt
  2012.05135}}].

\bibitem{Echevarria:2020hpy}
M.~G. Echevarria, Z.-B. Kang and J.~Terry, \emph{{Global analysis of the Sivers
  functions at NLO+NNLL in QCD}},
  \href{http://dx.doi.org/10.1007/JHEP01(2021)126}{\emph{JHEP} {\bf 01} (2021)
  126}, [\href{http://arxiv.org/abs/2009.10710}{{\tt 2009.10710}}].

\bibitem{Abramowicz:2015mha}
{\scshape H1, ZEUS} collaboration, H.~Abramowicz et~al., \emph{{Combination of
  measurements of inclusive deep inelastic ${e^{\pm }p}$ scattering cross
  sections and QCD analysis of HERA data}},
  \href{http://dx.doi.org/10.1140/epjc/s10052-015-3710-4}{\emph{Eur. Phys. J.}
  {\bf C75} (2015) 580}, [\href{http://arxiv.org/abs/1506.06042}{{\tt
  1506.06042}}].

\bibitem{Ball:2017nwa}
{\scshape NNPDF} collaboration, R.~D. Ball et~al., \emph{{Parton distributions
  from high-precision collider data}},
  \href{http://dx.doi.org/10.1140/epjc/s10052-017-5199-5}{\emph{Eur. Phys. J.}
  {\bf C77} (2017) 663}, [\href{http://arxiv.org/abs/1706.00428}{{\tt
  1706.00428}}].

\bibitem{Hou:2019efy}
T.-J. Hou et~al., \emph{{New CTEQ global analysis of quantum chromodynamics
  with high-precision data from the LHC}},
  \href{http://dx.doi.org/10.1103/PhysRevD.103.014013}{\emph{Phys. Rev. D} {\bf
  103} (2021) 014013}, [\href{http://arxiv.org/abs/1912.10053}{{\tt
  1912.10053}}].

\bibitem{Bailey:2020ooq}
S.~Bailey, T.~Cridge, L.~A. Harland-Lang, A.~D. Martin and R.~S. Thorne,
  \emph{{Parton distributions from LHC, HERA, Tevatron and fixed target data:
  MSHT20 PDFs}},
  \href{http://dx.doi.org/10.1140/epjc/s10052-021-09057-0}{\emph{Eur. Phys. J.
  C} {\bf 81} (2021) 341}, [\href{http://arxiv.org/abs/2012.04684}{{\tt
  2012.04684}}].

\bibitem{Hautmann:2020cyp}
F.~Hautmann, I.~Scimemi and A.~Vladimirov, \emph{{Non-perturbative
  contributions to vector-boson transverse momentum spectra in hadronic
  collisions}},
  \href{http://dx.doi.org/10.1016/j.physletb.2020.135478}{\emph{Phys. Lett. B}
  {\bf 806} (2020) 135478}, [\href{http://arxiv.org/abs/2002.12810}{{\tt
  2002.12810}}].

\bibitem{Collins:2011zzd}
J.~Collins, \emph{{Foundations of perturbative QCD}}.
\newblock Cambridge University Press, 2013.

\bibitem{Collins:1984kg}
J.~C. Collins, D.~E. Soper and G.~F. Sterman, \emph{{Transverse Momentum
  Distribution in Drell-Yan Pair and W and Z Boson Production}},
  \href{http://dx.doi.org/10.1016/0550-3213(85)90479-1}{\emph{Nucl. Phys.} {\bf
  B250} (1985) 199--224}.

\bibitem{Becher:2010tm}
T.~Becher and M.~Neubert, \emph{{{Drell-Yan} Production at Small $q_T$,
  Transverse Parton Distributions and the Collinear Anomaly}},
  \href{http://dx.doi.org/10.1140/epjc/s10052-011-1665-7}{\emph{Eur. Phys. J.}
  {\bf C71} (2011) 1665}, [\href{http://arxiv.org/abs/1007.4005}{{\tt
  1007.4005}}].

\bibitem{Echevarria:2011epo}
M.~G. Echevarria, A.~Idilbi and I.~Scimemi, \emph{{Factorization Theorem For
  Drell-Yan At Low $q_T$ And Transverse Momentum Distributions
  On-The-Light-Cone}},
  \href{http://dx.doi.org/10.1007/JHEP07(2012)002}{\emph{JHEP} {\bf 07} (2012)
  002}, [\href{http://arxiv.org/abs/1111.4996}{{\tt 1111.4996}}].

\bibitem{Chiu:2012ir}
J.-Y. Chiu, A.~Jain, D.~Neill and I.~Z. Rothstein, \emph{{A Formalism for the
  Systematic Treatment of Rapidity Logarithms in Quantum Field Theory}},
  \href{http://dx.doi.org/10.1007/JHEP05(2012)084}{\emph{JHEP} {\bf 05} (2012)
  084}, [\href{http://arxiv.org/abs/1202.0814}{{\tt 1202.0814}}].

\bibitem{Scimemi:2018xaf}
I.~Scimemi and A.~Vladimirov, \emph{{Systematic analysis of double-scale
  evolution}}, \href{http://dx.doi.org/10.1007/JHEP08(2018)003}{\emph{JHEP}
  {\bf 08} (2018) 003}, [\href{http://arxiv.org/abs/1803.11089}{{\tt
  1803.11089}}].

\bibitem{Gutierrez-Reyes:2019vbx}
D.~Gutierrez-Reyes, I.~Scimemi, W.~J. Waalewijn and L.~Zoppi, \emph{{Transverse
  momentum dependent distributions in $e^+e^-$ and semi-inclusive
  deep-inelastic scattering using jets}},
  \href{http://dx.doi.org/10.1007/JHEP10(2019)031}{\emph{JHEP} {\bf 10} (2019)
  031}, [\href{http://arxiv.org/abs/1904.04259}{{\tt 1904.04259}}].

\bibitem{Gutierrez-Reyes:2020ouu}
D.~Gutierrez-Reyes, S.~Leal-Gomez and I.~Scimemi, \emph{{W-boson production in
  TMD factorization}},
  \href{http://dx.doi.org/10.1140/epjc/s10052-021-09202-9}{\emph{Eur. Phys. J.
  C} {\bf 81} (2021) 418}, [\href{http://arxiv.org/abs/2011.05351}{{\tt
  2011.05351}}].

\bibitem{Moos:2020wvd}
V.~Moos and A.~Vladimirov, \emph{{Calculation of transverse momentum dependent
  distributions beyond the leading power}},
  \href{http://dx.doi.org/10.1007/JHEP12(2020)145}{\emph{JHEP} {\bf 12} (2020)
  145}, [\href{http://arxiv.org/abs/2008.01744}{{\tt 2008.01744}}].

\bibitem{Vladimirov:2021hdn}
A.~Vladimirov, V.~Moos and I.~Scimemi, \emph{{Transverse momentum dependent
  operator expansion at next-to-leading power}},
  \href{http://arxiv.org/abs/2109.09771}{{\tt 2109.09771}}.

\bibitem{Collins:1981uk}
J.~C. Collins and D.~E. Soper, \emph{{Back-To-Back Jets in QCD}},
  \href{http://dx.doi.org/10.1016/0550-3213(81)90339-4}{\emph{Nucl. Phys.} {\bf
  B193} (1981) 381}.

\bibitem{Collins:1981va}
J.~C. Collins and D.~E. Soper, \emph{{Back-To-Back Jets: Fourier Transform from
  B to K-Transverse}},
  \href{http://dx.doi.org/10.1016/0550-3213(82)90453-9}{\emph{Nucl. Phys.} {\bf
  B197} (1982) 446--476}.

\bibitem{Echevarria:2016scs}
M.~G. Echevarria, I.~Scimemi and A.~Vladimirov, \emph{{Unpolarized Transverse
  Momentum Dependent Parton Distribution and Fragmentation Functions at
  next-to-next-to-leading order}},
  \href{http://dx.doi.org/10.1007/JHEP09(2016)004}{\emph{JHEP} {\bf 09} (2016)
  004}, [\href{http://arxiv.org/abs/1604.07869}{{\tt 1604.07869}}].

\bibitem{Luo:2020epw}
M.-x. Luo, T.-Z. Yang, H.~X. Zhu and Y.~J. Zhu, \emph{{Unpolarized quark and
  gluon TMD PDFs and FFs at N$^{3}$LO}},
  \href{http://dx.doi.org/10.1007/JHEP06(2021)115}{\emph{JHEP} {\bf 06} (2021)
  115}, [\href{http://arxiv.org/abs/2012.03256}{{\tt 2012.03256}}].

\bibitem{Luo:2019szz}
M.-x. Luo, T.-Z. Yang, H.~X. Zhu and Y.~J. Zhu, \emph{{Quark Transverse Parton
  Distribution at the Next-to-Next-to-Next-to-Leading Order}},
  \href{http://dx.doi.org/10.1103/PhysRevLett.124.092001}{\emph{Phys. Rev.
  Lett.} {\bf 124} (2020) 092001}, [\href{http://arxiv.org/abs/1912.05778}{{\tt
  1912.05778}}].

\bibitem{Ebert:2020yqt}
M.~A. Ebert, B.~Mistlberger and G.~Vita, \emph{{Transverse momentum dependent
  PDFs at N$^3$LO}},
  \href{http://dx.doi.org/10.1007/JHEP09(2020)146}{\emph{JHEP} {\bf 09} (2020)
  146}, [\href{http://arxiv.org/abs/2006.05329}{{\tt 2006.05329}}].

\bibitem{Echevarria:2012pw}
M.~G. Echevarria, A.~Idilbi, A.~Schafer and I.~Scimemi,
  \emph{{Model-Independent Evolution of Transverse Momentum Dependent
  Distribution Functions (TMDs) at NNLL}},
  \href{http://dx.doi.org/10.1140/epjc/s10052-013-2636-y}{\emph{Eur. Phys. J.}
  {\bf C73} (2013) 2636}, [\href{http://arxiv.org/abs/1208.1281}{{\tt
  1208.1281}}].

\bibitem{Echevarria:2015byo}
M.~G. Echevarria, I.~Scimemi and A.~Vladimirov, \emph{{Universal transverse
  momentum dependent soft function at NNLO}},
  \href{http://dx.doi.org/10.1103/PhysRevD.93.054004}{\emph{Phys. Rev.} {\bf
  D93} (2016) 054004}, [\href{http://arxiv.org/abs/1511.05590}{{\tt
  1511.05590}}].

\bibitem{Vladimirov:2016dll}
A.~A. Vladimirov, \emph{{Soft-/rapidity- anomalous dimensions correspondence}},
  \href{http://dx.doi.org/10.1103/PhysRevLett.118.062001}{\emph{Phys. Rev.
  Lett.} {\bf 118} (2017) 062001}, [\href{http://arxiv.org/abs/1610.05791}{{\tt
  1610.05791}}].

\bibitem{Vladimirov:2020umg}
A.~A. Vladimirov, \emph{{Self-contained definition of Collins-Soper kernel}},
  \href{http://arxiv.org/abs/2003.02288}{{\tt 2003.02288}}.

\bibitem{Collins:2014jpa}
J.~Collins and T.~Rogers, \emph{{Understanding the large-distance behavior of
  transverse-momentum-dependent parton densities and the Collins-Soper
  evolution kernel}},
  \href{http://dx.doi.org/10.1103/PhysRevD.91.074020}{\emph{Phys. Rev.} {\bf
  D91} (2015) 074020}, [\href{http://arxiv.org/abs/1412.3820}{{\tt
  1412.3820}}].

\bibitem{LatticeParton:2020uhz}
{\scshape Lattice Parton} collaboration, Q.-A. Zhang et~al., \emph{{Lattice QCD
  Calculations of Transverse-Momentum-Dependent Soft Function through
  Large-Momentum Effective Theory}},
  \href{http://dx.doi.org/10.1103/PhysRevLett.125.192001}{\emph{Phys. Rev.
  Lett.} {\bf 125} (2020) 192001}, [\href{http://arxiv.org/abs/2005.14572}{{\tt
  2005.14572}}].

\bibitem{Schlemmer:2021aij}
M.~Schlemmer, A.~Vladimirov, C.~Zimmermann, M.~Engelhardt and A.~Sch\"afer,
  \emph{{Determination of the Collins-Soper Kernel from Lattice QCD}},
  \href{http://dx.doi.org/10.1007/JHEP08(2021)004}{\emph{JHEP} {\bf 08} (2021)
  004}, [\href{http://arxiv.org/abs/2103.16991}{{\tt 2103.16991}}].

\bibitem{Shanahan:2021tst}
P.~Shanahan, M.~Wagman and Y.~Zhao, \emph{{Lattice QCD calculation of the
  Collins-Soper kernel from quasi TMDPDFs}},
  \href{http://arxiv.org/abs/2107.11930}{{\tt 2107.11930}}.

\bibitem{Hautmann:2021ovt}
F.~Hautmann, I.~Scimemi and A.~Vladimirov, \emph{{Determination of the rapidity
  evolution kernel from Drell-Yan data at low transverse momenta}},  in
  \emph{{28th International Workshop on Deep Inelastic Scattering and Related
  Subjects}}, 9, 2021.
\newblock \href{http://arxiv.org/abs/2109.12051}{{\tt 2109.12051}}.

\bibitem{Konychev:2005iy}
A.~V. Konychev and P.~M. Nadolsky, \emph{{Universality of the
  Collins-Soper-Sterman nonperturbative function in gauge boson production}},
  \href{http://dx.doi.org/10.1016/j.physletb.2005.12.063}{\emph{Phys. Lett. B}
  {\bf 633} (2006) 710--714}, [\href{http://arxiv.org/abs/hep-ph/0506225}{{\tt
  hep-ph/0506225}}].

\bibitem{Landry:2002ix}
F.~Landry, R.~Brock, P.~M. Nadolsky and C.~P. Yuan, \emph{{Tevatron Run-1 $Z$
  boson data and Collins-Soper-Sterman resummation formalism}},
  \href{http://dx.doi.org/10.1103/PhysRevD.67.073016}{\emph{Phys. Rev.} {\bf
  D67} (2003) 073016}, [\href{http://arxiv.org/abs/hep-ph/0212159}{{\tt
  hep-ph/0212159}}].

\bibitem{Landry:1999an}
F.~Landry, R.~Brock, G.~Ladinsky and C.~P. Yuan, \emph{{New fits for the
  nonperturbative parameters in the CSS resummation formalism}},
  \href{http://dx.doi.org/10.1103/PhysRevD.63.013004}{\emph{Phys. Rev.} {\bf
  D63} (2001) 013004}, [\href{http://arxiv.org/abs/hep-ph/9905391}{{\tt
  hep-ph/9905391}}].

\bibitem{Hautmann:2007cx}
F.~Hautmann and D.~E. Soper, \emph{{Parton distribution function for quarks in
  an s-channel approach}},
  \href{http://dx.doi.org/10.1103/PhysRevD.75.074020}{\emph{Phys. Rev. D} {\bf
  75} (2007) 074020}, [\href{http://arxiv.org/abs/hep-ph/0702077}{{\tt
  hep-ph/0702077}}].

\bibitem{Hautmann:2000pw}
F.~Hautmann and D.~E. Soper, \emph{{Color transparency in deeply inelastic
  diffraction}},
  \href{http://dx.doi.org/10.1103/PhysRevD.63.011501}{\emph{Phys. Rev. D} {\bf
  63} (2001) 011501}, [\href{http://arxiv.org/abs/hep-ph/0008224}{{\tt
  hep-ph/0008224}}].

\bibitem{Hautmann:1999ui}
F.~Hautmann, Z.~Kunszt and D.~E. Soper, \emph{{Hard scattering factorization
  and light cone Hamiltonian approach to diffractive processes}},
  \href{http://dx.doi.org/10.1016/S0550-3213(99)00568-4}{\emph{Nucl. Phys. B}
  {\bf 563} (1999) 153--199}, [\href{http://arxiv.org/abs/hep-ph/9906284}{{\tt
  hep-ph/9906284}}].

\bibitem{Guzzi:2013aja}
M.~Guzzi, P.~M. Nadolsky and B.~Wang, \emph{{Nonperturbative contributions to a
  resummed leptonic angular distribution in inclusive neutral vector boson
  production}}, \href{http://dx.doi.org/10.1103/PhysRevD.90.014030}{\emph{Phys.
  Rev.} {\bf D90} (2014) 014030}, [\href{http://arxiv.org/abs/1309.1393}{{\tt
  1309.1393}}].

\bibitem{Schweitzer:2010tt}
P.~Schweitzer, T.~Teckentrup and A.~Metz, \emph{{Intrinsic transverse parton
  momenta in deeply inelastic reactions}},
  \href{http://dx.doi.org/10.1103/PhysRevD.81.094019}{\emph{Phys. Rev.} {\bf
  D81} (2010) 094019}, [\href{http://arxiv.org/abs/1003.2190}{{\tt
  1003.2190}}].

\bibitem{Schweitzer:2012hh}
P.~Schweitzer, M.~Strikman and C.~Weiss, \emph{{Intrinsic transverse momentum
  and parton correlations from dynamical chiral symmetry breaking}},
  \href{http://dx.doi.org/10.1007/JHEP01(2013)163}{\emph{JHEP} {\bf 01} (2013)
  163}, [\href{http://arxiv.org/abs/1210.1267}{{\tt 1210.1267}}].

\bibitem{Scimemi:2016ffw}
I.~Scimemi and A.~Vladimirov, \emph{{Power corrections and renormalons in
  Transverse Momentum Distributions}},
  \href{http://dx.doi.org/10.1007/JHEP03(2017)002}{\emph{JHEP} {\bf 03} (2017)
  002}, [\href{http://arxiv.org/abs/1609.06047}{{\tt 1609.06047}}].

\bibitem{web}
``\texttt{artemide} \text{web-page, https://teorica.fis.ucm.es/artemide/.} \\
  \texttt{artemide} \text{repository,
  https://github.com/VladimirovAlexey/artemide-public.} \text{The W-boson case
  is in https://github.com/SergioLealGomezTMD/artemide-development}.''

\bibitem{Buckley:2014ana}
A.~Buckley, J.~Ferrando, S.~Lloyd, K.~Noerdstrom, B.~Page, M.~Ruefenacht
  et~al., \emph{{LHAPDF6: parton density access in the LHC precision era}},
  \href{http://dx.doi.org/10.1140/epjc/s10052-015-3318-8}{\emph{Eur. Phys. J.}
  {\bf C75} (2015) 132}, [\href{http://arxiv.org/abs/1412.7420}{{\tt
  1412.7420}}].

\bibitem{iminuit}
H.~Dembinski and P.~O. et~al., \emph{scikit-hep/iminuit}, .

\bibitem{BermudezMartinez:2019anj}
A.~Bermudez~Martinez et~al., \emph{{Production of Z-bosons in the parton
  branching method}},
  \href{http://dx.doi.org/10.1103/PhysRevD.100.074027}{\emph{Phys. Rev. D} {\bf
  100} (2019) 074027}, [\href{http://arxiv.org/abs/1906.00919}{{\tt
  1906.00919}}].

\bibitem{BermudezMartinez:2020tys}
A.~Bermudez~Martinez et~al., \emph{{The transverse momentum spectrum of low
  mass Drell\textendash{}Yan production at next-to-leading order in the parton
  branching method}},
  \href{http://dx.doi.org/10.1140/epjc/s10052-020-8136-y}{\emph{Eur. Phys. J.
  C} {\bf 80} (2020) 598}, [\href{http://arxiv.org/abs/2001.06488}{{\tt
  2001.06488}}].

\bibitem{Collins:2000gd}
J.~C. Collins and F.~Hautmann, \emph{{Soft gluons and gauge invariant
  subtractions in NLO parton shower Monte Carlo event generators}},
  \href{http://dx.doi.org/10.1088/1126-6708/2001/03/016}{\emph{JHEP} {\bf 03}
  (2001) 016}, [\href{http://arxiv.org/abs/hep-ph/0009286}{{\tt
  hep-ph/0009286}}].

\bibitem{Ito:1980ev}
A.~S. Ito et~al., \emph{{Measurement of the Continuum of Dimuons Produced in
  High-Energy Proton - Nucleus Collisions}},
  \href{http://dx.doi.org/10.1103/PhysRevD.23.604}{\emph{Phys. Rev.} {\bf D23}
  (1981) 604--633}.

\bibitem{Moreno:1990sf}
G.~Moreno et~al., \emph{{Dimuon production in proton - copper collisions at
  $\sqrt{s}$ = 38.8-GeV}},
  \href{http://dx.doi.org/10.1103/PhysRevD.43.2815}{\emph{Phys. Rev.} {\bf D43}
  (1991) 2815--2836}.

\bibitem{McGaughey:1994dx}
{\scshape E772} collaboration, P.~L. McGaughey et~al., \emph{{Cross-sections
  for the production of high mass muon pairs from 800-GeV proton bombardment of
  H-2}}, \href{http://dx.doi.org/10.1103/PhysRevD.50.3038,
  10.1103/PhysRevD.60.119903}{\emph{Phys. Rev.} {\bf D50} (1994) 3038--3045}.

\bibitem{Aidala:2018ajl}
{\scshape PHENIX} collaboration, C.~Aidala et~al., \emph{{Measurements of
  $\mu\mu$ pairs from open heavy flavor and Drell-Yan in $p+p$ collisions at
  $\sqrt{s}=200$ GeV}}, {\emph{Submitted to: Phys. Rev. D} (2018) },
  [\href{http://arxiv.org/abs/1805.02448}{{\tt 1805.02448}}].

\bibitem{Affolder:1999jh}
{\scshape CDF} collaboration, T.~Affolder et~al., \emph{{The transverse
  momentum and total cross section of $e^+e^-$ pairs in the $Z$ boson region
  from $p\bar{p}$ collisions at $\sqrt{s} = 1.8$ TeV}},
  \href{http://dx.doi.org/10.1103/PhysRevLett.84.845}{\emph{Phys. Rev. Lett.}
  {\bf 84} (2000) 845--850}, [\href{http://arxiv.org/abs/hep-ex/0001021}{{\tt
  hep-ex/0001021}}].

\bibitem{Aaltonen:2012fi}
{\scshape CDF} collaboration, T.~Aaltonen et~al., \emph{{Transverse momentum
  cross section of $e^+e^-$ pairs in the $Z$-boson region from $p\bar{p}$
  collisions at $\sqrt{s}=1.96$ TeV}},
  \href{http://dx.doi.org/10.1103/PhysRevD.86.052010}{\emph{Phys. Rev.} {\bf
  D86} (2012) 052010}, [\href{http://arxiv.org/abs/1207.7138}{{\tt
  1207.7138}}].

\bibitem{Abbott:1999wk}
{\scshape D0} collaboration, B.~Abbott et~al., \emph{{Measurement of the
  inclusive differential cross section for $Z$ bosons as a function of
  transverse momentum in $\bar{p}p$ collisions at $\sqrt{s} = 1.8$ TeV}},
  \href{http://dx.doi.org/10.1103/PhysRevD.61.032004}{\emph{Phys. Rev.} {\bf
  D61} (2000) 032004}, [\href{http://arxiv.org/abs/hep-ex/9907009}{{\tt
  hep-ex/9907009}}].

\bibitem{Abazov:2007ac}
{\scshape D0} collaboration, V.~M. Abazov et~al., \emph{{Measurement of the
  shape of the boson transverse momentum distribution in $p \bar{p} \to Z /
  \gamma^{*} \to e^+ e^- + X$ events produced at $\sqrt{s}$=1.96-TeV}},
  \href{http://dx.doi.org/10.1103/PhysRevLett.100.102002}{\emph{Phys. Rev.
  Lett.} {\bf 100} (2008) 102002}, [\href{http://arxiv.org/abs/0712.0803}{{\tt
  0712.0803}}].

\bibitem{Abazov:2010kn}
{\scshape D0} collaboration, V.~M. Abazov et~al., \emph{{Measurement of the
  normalized $Z/\gamma^* -> \mu^+\mu^-$ transverse momentum distribution in
  $p\bar{p}$ collisions at $\sqrt{s}=1.96$ TeV}},
  \href{http://dx.doi.org/10.1016/j.physletb.2010.09.012}{\emph{Phys. Lett.}
  {\bf B693} (2010) 522--530}, [\href{http://arxiv.org/abs/1006.0618}{{\tt
  1006.0618}}].

\bibitem{Aad:2014xaa}
{\scshape ATLAS} collaboration, G.~Aad et~al., \emph{{Measurement of the
  $Z/\gamma^*$ boson transverse momentum distribution in $pp$ collisions at
  $\sqrt{s}$ = 7 TeV with the ATLAS detector}},
  \href{http://dx.doi.org/10.1007/JHEP09(2014)145}{\emph{JHEP} {\bf 09} (2014)
  145}, [\href{http://arxiv.org/abs/1406.3660}{{\tt 1406.3660}}].

\bibitem{Aad:2015auj}
{\scshape ATLAS} collaboration, G.~Aad et~al., \emph{{Measurement of the
  transverse momentum and $\phi ^*_{\eta }$ distributions of Drell-Yan lepton
  pairs in proton-proton collisions at $\sqrt{s}=8$ TeV with the ATLAS
  detector}},
  \href{http://dx.doi.org/10.1140/epjc/s10052-016-4070-4}{\emph{Eur. Phys. J.}
  {\bf C76} (2016) 291}, [\href{http://arxiv.org/abs/1512.02192}{{\tt
  1512.02192}}].

\bibitem{Chatrchyan:2011wt}
{\scshape CMS} collaboration, S.~Chatrchyan et~al., \emph{{Measurement of the
  Rapidity and Transverse Momentum Distributions of $Z$ Bosons in $pp$
  Collisions at $\sqrt{s}=7$ TeV}},
  \href{http://dx.doi.org/10.1103/PhysRevD.85.032002}{\emph{Phys. Rev.} {\bf
  D85} (2012) 032002}, [\href{http://arxiv.org/abs/1110.4973}{{\tt
  1110.4973}}].

\bibitem{Khachatryan:2016nbe}
{\scshape CMS} collaboration, V.~Khachatryan et~al., \emph{{Measurement of the
  transverse momentum spectra of weak vector bosons produced in proton-proton
  collisions at $ \sqrt{s}=8 $ TeV}},
  \href{http://dx.doi.org/10.1007/JHEP02(2017)096}{\emph{JHEP} {\bf 02} (2017)
  096}, [\href{http://arxiv.org/abs/1606.05864}{{\tt 1606.05864}}].

\bibitem{CMS:2019raw}
{\scshape CMS} collaboration, A.~M. Sirunyan et~al., \emph{{Measurements of
  differential Z boson production cross sections in proton-proton collisions at
  $ \sqrt{s} $ = 13 TeV}},
  \href{http://dx.doi.org/10.1007/JHEP12(2019)061}{\emph{JHEP} {\bf 12} (2019)
  061}, [\href{http://arxiv.org/abs/1909.04133}{{\tt 1909.04133}}].

\bibitem{Aaij:2015gna}
{\scshape LHCb} collaboration, R.~Aaij et~al., \emph{{Measurement of the
  forward $Z$ boson production cross-section in $pp$ collisions at $\sqrt{s}=7$
  TeV}}, \href{http://dx.doi.org/10.1007/JHEP08(2015)039}{\emph{JHEP} {\bf 08}
  (2015) 039}, [\href{http://arxiv.org/abs/1505.07024}{{\tt 1505.07024}}].

\bibitem{Aaij:2015zlq}
{\scshape LHCb} collaboration, R.~Aaij et~al., \emph{{Measurement of forward W
  and Z boson production in $pp$ collisions at $ \sqrt{s}=8 $ TeV}},
  \href{http://dx.doi.org/10.1007/JHEP01(2016)155}{\emph{JHEP} {\bf 01} (2016)
  155}, [\href{http://arxiv.org/abs/1511.08039}{{\tt 1511.08039}}].

\bibitem{Aaij:2016mgv}
{\scshape LHCb} collaboration, R.~Aaij et~al., \emph{{Measurement of the
  forward Z boson production cross-section in pp collisions at $\sqrt{s} = 13$
  TeV}}, \href{http://dx.doi.org/10.1007/JHEP09(2016)136}{\emph{JHEP} {\bf 09}
  (2016) 136}, [\href{http://arxiv.org/abs/1607.06495}{{\tt 1607.06495}}].

\bibitem{Ball:2008by}
{\scshape NNPDF} collaboration, R.~D. Ball, L.~Del~Debbio, S.~Forte,
  A.~Guffanti, J.~I. Latorre, A.~Piccione et~al., \emph{{A Determination of
  parton distributions with faithful uncertainty estimation}},
  \href{http://dx.doi.org/10.1016/j.nuclphysb.2008.09.037,
  10.1016/j.nuclphysb.2009.02.027}{\emph{Nucl. Phys.} {\bf B809} (2009) 1--63},
  [\href{http://arxiv.org/abs/0808.1231}{{\tt 0808.1231}}].

\bibitem{Ball:2012wy}
R.~D. Ball et~al., \emph{{Parton Distribution Benchmarking with LHC Data}},
  \href{http://dx.doi.org/10.1007/JHEP04(2013)125}{\emph{JHEP} {\bf 04} (2013)
  125}, [\href{http://arxiv.org/abs/1211.5142}{{\tt 1211.5142}}].

\bibitem{Hou:2016sho}
T.-J. Hou et~al., \emph{{Reconstruction of Monte Carlo replicas from Hessian
  parton distributions}},
  \href{http://dx.doi.org/10.1007/JHEP03(2017)099}{\emph{JHEP} {\bf 03} (2017)
  099}, [\href{http://arxiv.org/abs/1607.06066}{{\tt 1607.06066}}].

\bibitem{Bacchetta:2022awv}
A.~Bacchetta, V.~Bertone, C.~Bissolotti, G.~Bozzi, M.~Cerutti, F.~Piacenza
  et~al., \emph{{Unpolarized Transverse Momentum Distributions from a global
  fit of Drell-Yan and Semi-Inclusive Deep-Inelastic Scattering data}},
  \href{http://arxiv.org/abs/2206.07598}{{\tt 2206.07598}}.

\bibitem{Musch:2011er}
B.~U. Musch, P.~Hagler, M.~Engelhardt, J.~W. Negele and A.~Schafer,
  \emph{{Sivers and Boer-Mulders observables from lattice QCD}},
  \href{http://dx.doi.org/10.1103/PhysRevD.85.094510}{\emph{Phys. Rev. D} {\bf
  85} (2012) 094510}, [\href{http://arxiv.org/abs/1111.4249}{{\tt 1111.4249}}].

\bibitem{Abe:1991rk}
{\scshape CDF} collaboration, F.~Abe et~al., \emph{{Measurement of the W P(T)
  distribution in $\bar{p}p$ collisions at $\sqrt{s} = 1.8$ TeV}},
  \href{http://dx.doi.org/10.1103/PhysRevLett.66.2951}{\emph{Phys. Rev. Lett.}
  {\bf 66} (1991) 2951--2955}.

\bibitem{Abbott:1998jy}
{\scshape D0} collaboration, B.~Abbott et~al., \emph{{Measurement of the shape
  of the transverse momentum distribution of $W$ bosons produced in $p\bar{p}$
  collisions at $\sqrt{s} = 1.8$ TeV}},
  \href{http://dx.doi.org/10.1103/PhysRevLett.80.5498}{\emph{Phys. Rev. Lett.}
  {\bf 80} (1998) 5498--5503}, [\href{http://arxiv.org/abs/hep-ex/9803003}{{\tt
  hep-ex/9803003}}].

\bibitem{Aad:2011fp}
{\scshape ATLAS} collaboration, G.~Aad et~al., \emph{{Measurement of the
  Transverse Momentum Distribution of $W$ Bosons in $pp$ Collisions at
  $\sqrt{s}=7$ TeV with the ATLAS Detector}},
  \href{http://dx.doi.org/10.1103/PhysRevD.85.012005}{\emph{Phys. Rev. D} {\bf
  85} (2012) 012005}, [\href{http://arxiv.org/abs/1108.6308}{{\tt 1108.6308}}].

\bibitem{Signori:2013mda}
A.~Signori, A.~Bacchetta, M.~Radici and G.~Schnell, \emph{{Investigations into
  the flavor dependence of partonic transverse momentum}},
  \href{http://dx.doi.org/10.1007/JHEP11(2013)194}{\emph{JHEP} {\bf 11} (2013)
  194}, [\href{http://arxiv.org/abs/1309.3507}{{\tt 1309.3507}}].

\bibitem{Bacchetta:2018lna}
A.~Bacchetta, G.~Bozzi, M.~Radici, M.~Ritzmann and A.~Signori, \emph{{Effect of
  Flavor-Dependent Partonic Transverse Momentum on the Determination of the $W$
  Boson Mass in Hadronic Collisions}},
  \href{http://dx.doi.org/10.1016/j.physletb.2018.11.002}{\emph{Phys. Lett. B}
  {\bf 788} (2019) 542--545}, [\href{http://arxiv.org/abs/1807.02101}{{\tt
  1807.02101}}].

\bibitem{Signori:2016lvd}
A.~Signori, \emph{{Flavor and Evolution Effects in TMD Phenomenology}:
  {Manifestation of Hadron Structure in High-Energy Scattering Processes}}.
\newblock PhD thesis, Vrije U., Amsterdam, 2016.

\bibitem{Harland-Lang:2014zoa}
L.~A. Harland-Lang, A.~D. Martin, P.~Motylinski and R.~S. Thorne, \emph{{Parton
  distributions in the LHC era: MMHT 2014 PDFs}},
  \href{http://dx.doi.org/10.1140/epjc/s10052-015-3397-6}{\emph{Eur. Phys. J.}
  {\bf C75} (2015) 204}, [\href{http://arxiv.org/abs/1412.3989}{{\tt
  1412.3989}}].

\bibitem{Dulat:2015mca}
S.~Dulat, T.-J. Hou, J.~Gao, M.~Guzzi, J.~Huston, P.~Nadolsky et~al.,
  \emph{{New parton distribution functions from a global analysis of quantum
  chromodynamics}},
  \href{http://dx.doi.org/10.1103/PhysRevD.93.033006}{\emph{Phys. Rev.} {\bf
  D93} (2016) 033006}, [\href{http://arxiv.org/abs/1506.07443}{{\tt
  1506.07443}}].

\bibitem{Butterworth:2015oua}
J.~Butterworth et~al., \emph{{PDF4LHC recommendations for LHC Run II}},
  \href{http://dx.doi.org/10.1088/0954-3899/43/2/023001}{\emph{J. Phys.} {\bf
  G43} (2016) 023001}, [\href{http://arxiv.org/abs/1510.03865}{{\tt
  1510.03865}}].

\bibitem{Accardi:2016qay}
A.~Accardi, L.~T. Brady, W.~Melnitchouk, J.~F. Owens and N.~Sato,
  \emph{{Constraints on large-$x$ parton distributions from new weak boson
  production and deep-inelastic scattering data}},
  \href{http://dx.doi.org/10.1103/PhysRevD.93.114017}{\emph{Phys. Rev. D} {\bf
  93} (2016) 114017}, [\href{http://arxiv.org/abs/1602.03154}{{\tt
  1602.03154}}].

\end{thebibliography}\endgroup
\end{document}